\documentclass[prb,aps,twocolumn,superscriptaddress,showkeys,groupaddress,longbibliography]{revtex4-1} 
\usepackage{amsmath}  
\usepackage{amsfonts} 
\usepackage{graphicx} 
\usepackage{tikz}
\usepackage[siunitx]{circuitikz} 
\usetikzlibrary{shapes,arrows}
\usetikzlibrary{decorations.pathmorphing}
\usepackage{siunitx}
\usepackage{wasysym}
\usepackage{booktabs}
\usepackage[utf8]{inputenc}
\usepackage{afterpage}

\usepackage[nooneline]{subfigure}

\begin{document}


\title{Optical quality assurance of GEM foils}

\author{T.~Hild\'en}
\email{timo.hilden@helsinki.fi}
\affiliation{Helsinki Institute of Physics, Helsinki, 00560, Finland}
\author{E.~Br\"ucken}
\affiliation{Helsinki Institute of Physics, Helsinki, 00560, Finland}
\author{J.~Heino}
\affiliation{Helsinki Institute of Physics, Helsinki, 00560, Finland}
\author{M.~Kalliokoski}
\affiliation{Helsinki Institute of Physics, Helsinki, 00560, Finland}
\author{A.~Karadzhinova}
\affiliation{Helsinki Institute of Physics, Helsinki, 00560, Finland}
\author{R.~Lauhakangas}
\affiliation{Helsinki Institute of Physics, Helsinki, 00560, Finland}
\author{E.~Tuominen}
\affiliation{Helsinki Institute of Physics, Helsinki, 00560, Finland}
\author{R.~Turpeinen}
\affiliation{Helsinki Institute of Physics, Helsinki, 00560, Finland}

\begin{abstract}
An analysis software was developed for the high aspect ratio optical scanning system in the Detector Laboratory of the University of Helsinki and the Helsinki Institute of Physics. The system is used e.g. in the quality assurance of the GEM-TPC detectors being developed for the beam diagnostics system of the SuperFRS at future FAIR facility. The software was tested by analyzing five CERN standard GEM foils scanned with the optical scanning system. The measurement uncertainty of the diameter of the GEM holes and the pitch of the hole pattern was found to be 0.5~$\mu$m and 0.3~$\mu$m, respectively. The software design and the performance are discussed. The correlation between the GEM hole size distribution and the corresponding gain variation was studied by comparing them against a detailed gain mapping of a foil and a set of six lower precision control measurements. It can be seen that a qualitative estimation of the behavior of the local variation in gain across the GEM foil can be made based on the measured sizes of the outer and inner holes.
\end{abstract}


\maketitle


\section{Introduction}

\begin{figure}[!h]\centering
\includegraphics[width=0.25\textwidth]{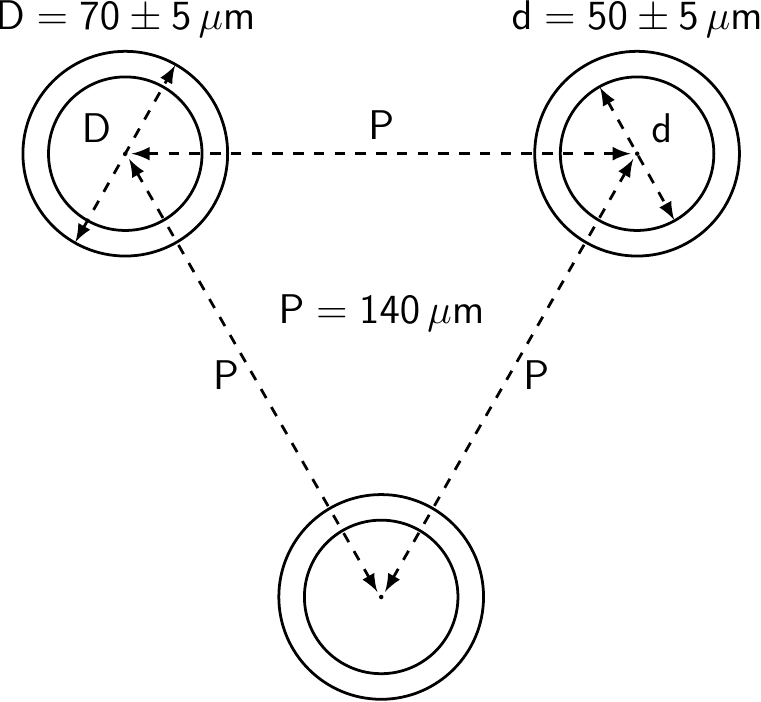}
\caption{A schematic illustration of GEM holes.\label{GEMschematic}}
\end{figure}

\begin{figure}[!h]\centering
\includegraphics[width=0.48\textwidth]{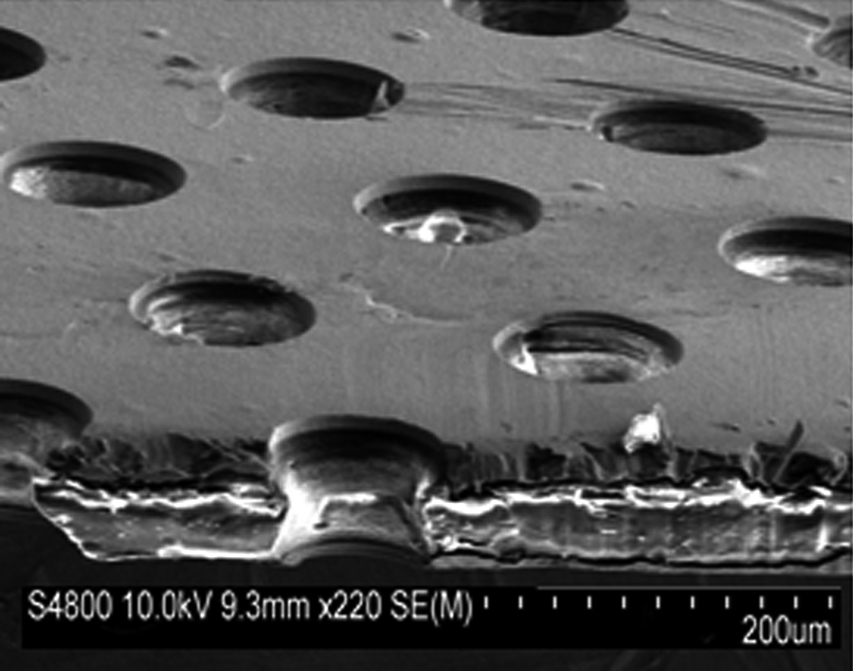}
\caption{A SEM image of the GEM hole profile.\label{SEMGEM}}
\end{figure}

GEM (Gaseous Electron Multiplier) detectors are a novel type of gas-filled radiation detectors~\citep{Sauli:1997qp} used in high energy physics (HEP) and in nuclear physics experiments. The basic detecting element of a GEM-detector is a thin (50~$\mu$m) polyimide foil with 5~$\mu$m copper coating on both sides. The foils have holes etched through them, with a typical pitch of 140~$\mu$m. The holes have an hour-glass like shape with an outer diameter of 70~$\mu$m and an inner diameter of 50~$\mu$m. Typically, a GEM foil with an area of 10 cm by 10 cm contains 600 000 holes. A schematic view of the GEM foil layout is shown in Figure~\ref{GEMschematic}, and a cutout SEM image of a foil in Figure~\ref{SEMGEM}. The electric field inside the holes can be controlled by applying voltage over the copper surfaces. Electron multiplication is achieved when the electric field inside a hole reaches the limit of avalanche formation for the gas, which is typically of the order of 10 kV/cm.

Primary electrons, freed from gas atoms by incident ionizing radiation, drift following the electric field into the holes, where they undergo avalanche multiplication. Secondary electrons and ions created in the avalanche continue to drift in the detector. The shape of the electric field near and inside the holes plays an important role in several key characteristics of GEM foils, such as electron transparency, ion feedback and rate dependent charge-up properties~\citep{Bachmann:1999xc, Sauli:2005zx, Alfonsi20126, Bachmann:2000az}. The field shape depends on the shape of the hole.

Local variation in the size or the shape of the holes can introduce variation in the operational characteristics of the GEM foil. Thus, in order to achieve uniform functionality over the active surface of the detector, the distribution and size of the holes should be as uniform as possible.

Since the manufacturing of the GEM foils is a complicated process, occasional defects in the form of areas with under- or over-etched holes can occur in the foils. The effects of etching defects to the performance of a GEM detector are not thoroughly understood, but observations of them causing discharges or even short circuits have been reported~\citep{5874011}. In addition to the defects resulting from the manufacturing process, the operation of the foils can be compromised by foreign matter as e.g. chemical residue, droplets of glue or small particles attached to the holes.

Recent development in the use of the GEM detectors has a trend towards detectors with larger active area and, especially in HEP experiments, larger mass productions of detectors~\citep{Ketzer:2013laa,Abbaneo:2013yua,Archilli:2010xb}. After the installation in a large HEP or nuclear physics experiment, access to the detectors is limited, thus, replacing broken detectors is difficult and can only take place during shutdowns that are usually made once a year. Therefore, the installed detectors must be of high quality and the development of quality assurance (QA) methods is crucial for the mass production of the GEM detectors for these experiments.

Several optical scanning setups for GEM foil characterization have been developed earlier, e.g.~\citep{Becker:2006yc,Simon_2007}. One of these, the large area optical scanning setup, was built in the Detector Laboratory of the University of Helsinki and the Helsinki Institute of Physics~\citep{Kalliokoski2012223,1748-0221-7-02-C02059}. It has mainly been applied in the quality assurance of the GEM-TPC detectors being developed for the beam diagnostics of the Super-FRS (Superconducting Fragment Separator) at the future FAIR (Facility for Antiproton and Ion Research) facility~\citep{Kalliokoski:2010zz} and for the ALICE TPC upgrade~\citep{Ketzer:2013laa} at CERN. The scanner setup consists of an xyz-table, a camera and software for controlling the table. However, trustworthy and versatile analysis software is a prerequisite for the usability of the setup.

Large area measurements of GEM foil characteristics are useful for foil manufacturers, too. The general tendency to increase the area of the GEM foils will also affect the production standards to maintain the homogeneity of the holes and to limit the amount of etching defects per unit of surface. Thus, reliable QA reflects to the yield of the manufacturing process. The scanning setup with the analysis software described in this paper is an excellent method of obtaining direct feedback for the tuning of the GEM foil manufacturing process.

\section{Analysis software}

In this work, the analysis software is implemented in Python which is an object oriented scripting language that is portable across different operating systems. In addition the Numpy~\citep{dubois.hinsen.hugunin-1996-cp} library is used for efficient operations on large arrays, and the OpenCV~\citep{Bradski:2000:OL} library for image processing.

\begin{figure}\centering
\includegraphics[width=0.48\textwidth]{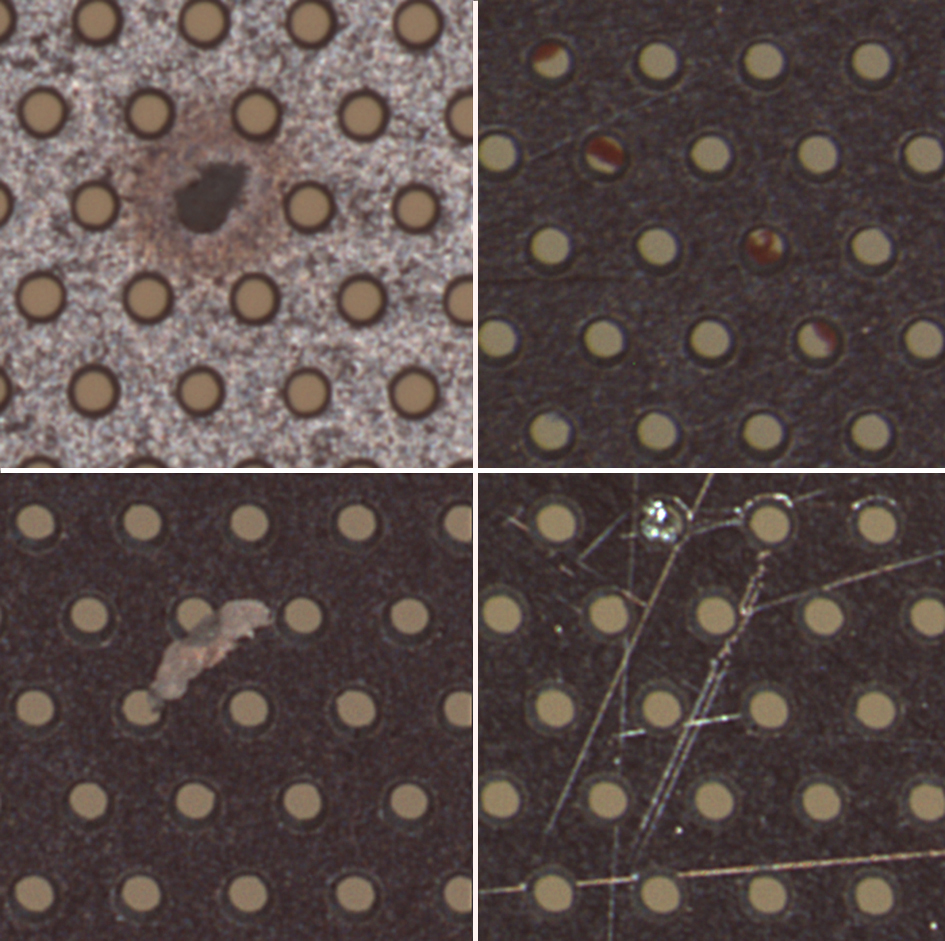}
\caption{Examples of a small etching defect (top left), dust (top right and bottom left) and scratches (bottom right). The image in the top left corner is taken of a matt foil, all the others are taken of a mirror-like foil.\label{GEMdefects}}
\end{figure}

The optical analysis of the images taken of the GEM foils was complicated by the fact that the surface reflectance of the copper surface varied strongly between individual foils. Images taken from a reflective, mirror-like foil were dark compared to the images of more diffusely reflecting matt foils. The top left image of Figure~\ref{GEMdefects} is taken from a matt foil, the three others are from a mirror foil. 

\begin{figure}\centering
\includegraphics[width=0.48\textwidth]{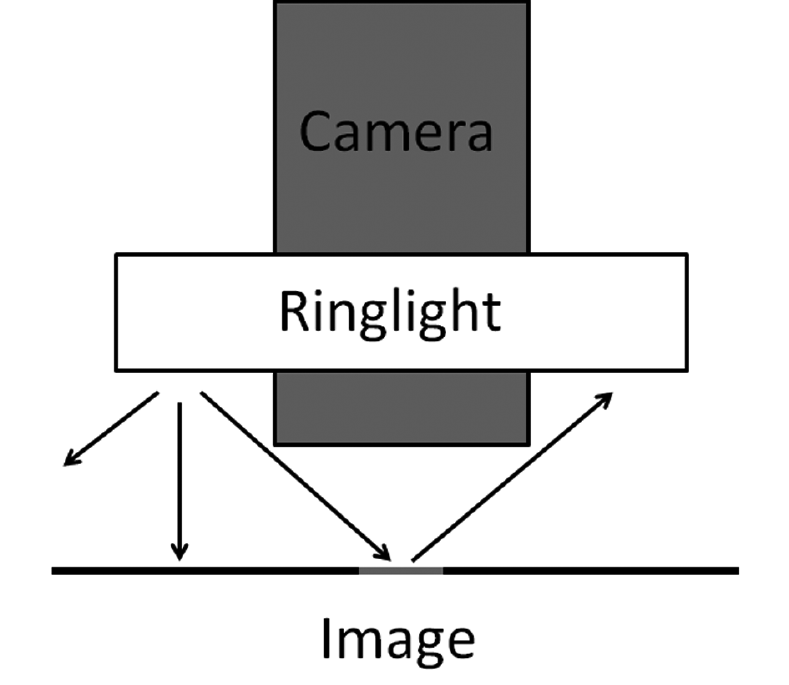}

\caption{An illustration of light reflection by the mirror foils.\label{lightreflection}}
\end{figure}

The mirror foils appear dark because the specular reflection of the foreground lighting could not reach the camera, as illustrated in Figure~\ref{lightreflection}. 

In addition to the variable reflectivity, the foil surface was often textured with scratches, dust and, occasionally, with oxidation or chemical residue from the manufacturing process. Example images of some of these are shown in Figure~\ref{GEMdefects}.

\begin{figure}\centering
\includegraphics[width=0.48\textwidth]{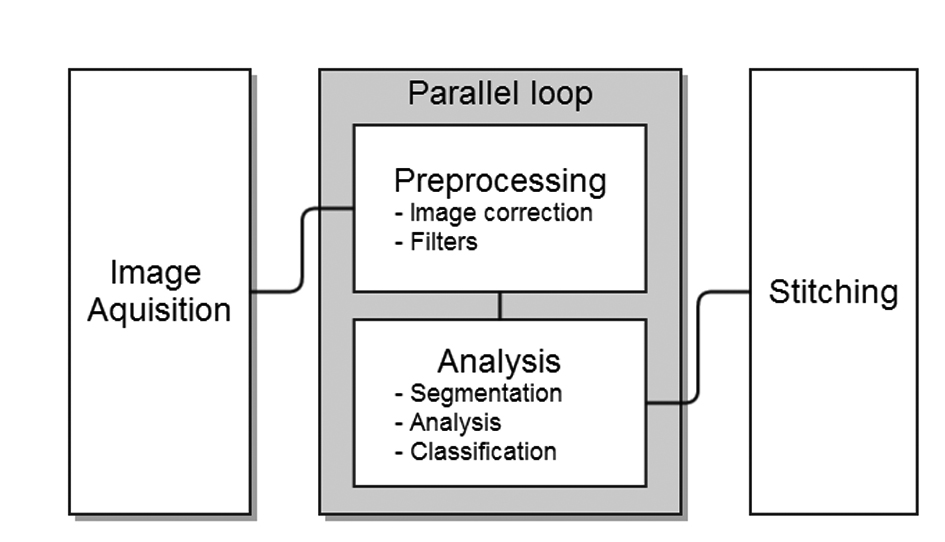}
\caption{A flowchart of the analysis process.\label{flowchart}}
\end{figure}

In order to extract the foil characteristics from the highly variable input images, the software was designed to be as generic as reasonably achievable in terms of image preprocessing, segmentation and data analysis. As a consequence, the software is independent of the image acquisition method. This requirement for the generic approach led to a design in which the image preprocessing algorithm, the segmentation and the object classification method are separated from each other and can be individually determined by a setup file tuned for each task.

The image analysis goes through several phases as illustrated in Figure~\ref{flowchart}. Each phase is independent and can be run in parallel, even with different computers if needed. The preprocessing and analysis phases are, however, implemented as a threaded parallel loop to minimize the number of file operations, and to optimize the scheduling of the tasks.

In the preprocessing phase, the images acquired in the previous phase are loaded into memory and optionally corrected for lens distortions. Then, the images are processed with a preprocessing algorithm to reduce noise and to make features of interest stand out for later stages. For this work the preprocessing algorithm consisted of a contrast enhancement filter. The filter was applied selectively to the image by a mask obtained by thresholding the image such that all but the foreground of the image was masked. The mask was then grown and blurred to include the neighborhood of the features to the filtered area. The purpose of the preprocessing algorithm was to make the outer holes and the defects detectable despite the highly contrasting texture of the copper foreground. Similar algorithm was used for the mirror foil to filter scratches and small bright spots from the image. 

In the analysis phase the image is segmented, or the features of interest are isolated from the processed image. Edge detection by the Canny algorithm~\citep{DBLP:journals/pami/Canny86a} was used for segmentation in this work. The edge contour and the underlying image is then analyzed to obtain the parameters of the features. These are the coordinates of the centroid and the size and coordinates of the bounding box of the feature, the six invariant Hu moments~\citep{1057692} of the contour, the major and minor axes of an ellipse fitted to the contour, as well as the angle between the ellipse major axis and the x-axis. The mean value of each color channel under the contour and under the convex hull of the contour is also recorded. This data is then fed into the classification algorithm. The features are either discarded or saved as one of the classes that are to be measured, such as ``inner hole'', ``outer hole'' or ``defect''. Classification can be made with predefined cuts of the data, or with a supervised learning method using a feed-forward neural network available in the OpenCV library. 

Classification by predetermined cuts is relatively straightforward, but a neural network classifier needs to be trained. To train the neural network, a set of sample images is chosen by the user. The samples, the analyzed data and the user defined classification are saved into the classifier. The neural network classifier handles the samples and the data, divides the samples into training and validation sets and trains the neural network. The saved set of samples can later be reviewed and features can be added, removed or switched into another class. The number of inputs and the size of the hidden layer can be modified and the new neural network will be trained using the samples.

The neural network classifier was used for classification in this work. The advantage of using the neural network classifier was the effective recognition of such features in which the contour of an outer hole was connected with a contour of a contrasting feature in the copper foreground or where the inner and outer hole contours had been connected. In such features the fitted ellipse would not follow the real boundary of the hole and the data would not represent the real size of the hole. Such features were discarded.

The input layer of the neural network was fed with all the data of a feature except the centroid coordinates. A single hidden layer was found sufficient between the input and the output layer. The output layer consisted of the four classifications, which were defined as `discarded', `inner hole', `outer hole' and `defect'. The `discarded' features were discarded from the analysis, the other classes of features were recorded. Small thumbnail images of features of the `defect' class were recorded for later inspection. The data of each image is saved on disk after analysis phase.

Finally, in the stitching phase the data is aligned with the template matching method. In the template matching method, the location of a smaller template image is found in relation to another, larger image. The template is matched on top of the base image, starting from the upper right corner. The squared difference of the pixel intensities between the images is calculated. Then the template is shifted by a pixel and new squared difference value is calculated. The process is repeated until the template has been moved through all possible locations in the base image. A map of these values creates a search surface for locating the template. The minimum value of the search surface corresponds to the offset of the template image. 
 
Accuracy of the movement of the scanning system between two neighboring images is of the order of 5 microns, well below the periodicity of the hole pattern on the GEM foils. This simplifies the stitching process considerably. Images are taken with 245~$\mu$m overlap between the neighboring images. The overlapping portion of each image - 140 pixel wide strips along each image edge - are saved with the data. When stitching, the overlapping image strip of one data file is taken as the template and matched against the other. 25 pixels, or about 44~$\mu$m, are stripped from each edge of the template image to provide a search surface of 50 pixels by 50 pixels. The result of the template matching is the offset of the stitched image in relation to the expected step size of the scanner.
 
\begin{figure}\centering
\includegraphics[width=0.48\textwidth]{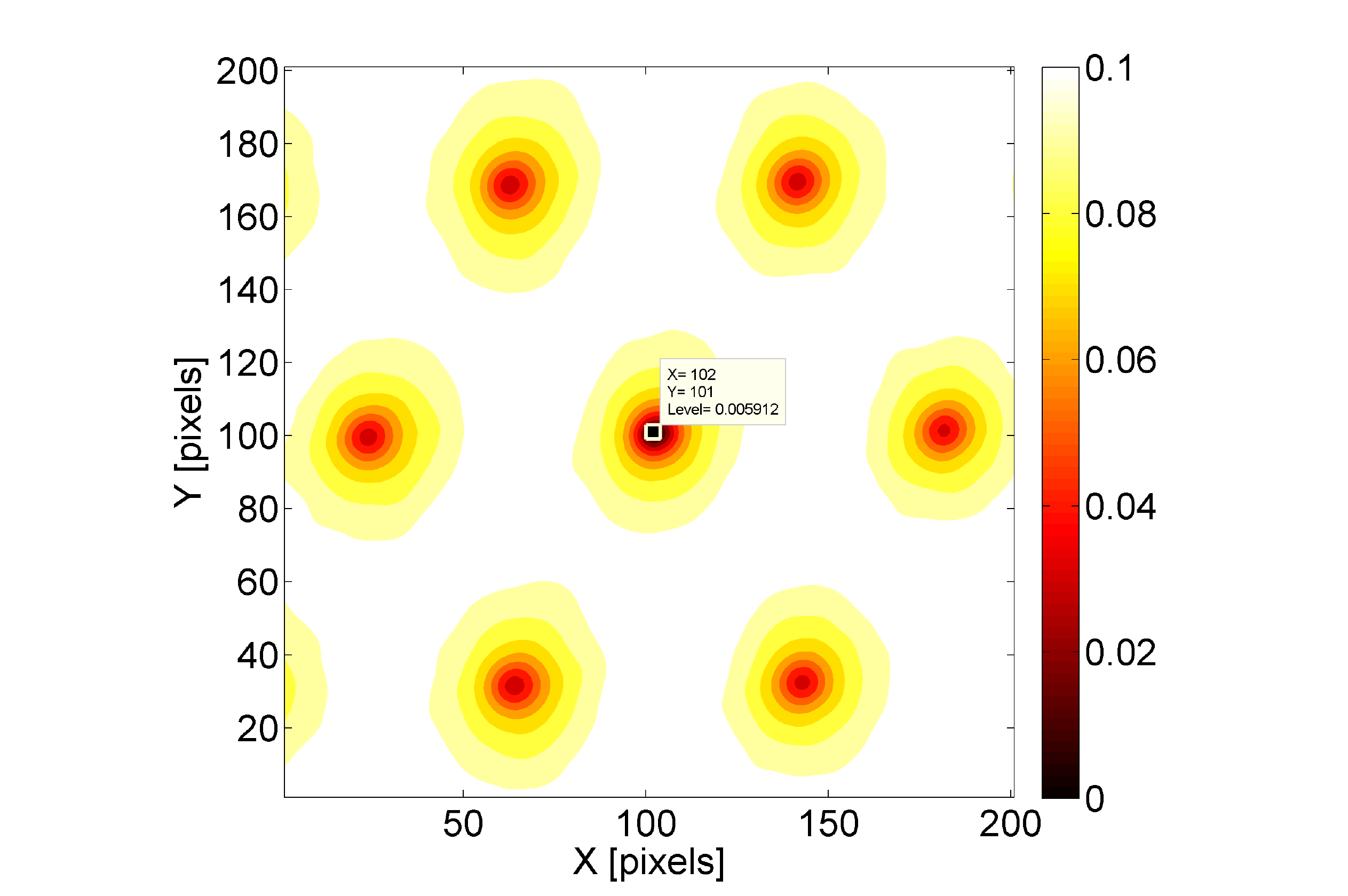}
\caption{The sum of the squared difference of pixel intensities of overlapping portions of two stitched images.\label{pixeldensities}}
\end{figure}

An example search surface of two horizontally neighboring images with 280 pixel overlap is shown in Figure~\ref{pixeldensities}, where the template size is 80 pixels by 2416 pixels. To emphasize the fact that the method can find the correct offset when crossing the periodicity of the hole pattern, the search surface was increased to 200 pixels by 200 pixels. The template image size for the stitching used in this work is 3458 pixels by 90 pixels in the horizontal edges and 90 pixels by 2566 pixels in the vertical edges, but only 90 pixels by 90 pixels in the diagonal directions.

A map of connecting images is created from the image coordinates and the data files are traversed through, one by one, according to the map. The lower left corner of the first data file is positioned at the origin of the coordinate system. All neighboring data files are stitched to its coordinates using the template matching method. The location data is saved with the data of each file. Overlapping features on the data files are found and only those on the sharper of the images are kept.

One of the stitched neighbors is then chosen and its neighbors are stitched relative to its coordinates. Location data of files that are stitched several times is defined as the average of the separate stitches. The search surface in the diagonal direction is very small compared to the search surface in side by side stitching. Diagonal stitches are given very small weight to minimize possible error. It is significant only, if there are no direct edge neighbors to stitch into - a situation which should not happen under normal running.

A substantial challenge in the microscopic imaging over large GEM foil surface area is the amount of data that is produced. The area of a single image taken by the scanning system is 6.1 mm by 4.6 mm, corresponding to 3488 by 2616 pixels, respectively. This takes roughly 26 MB in memory, and, once compressed, from 15 to 25 MB in the storage device. The full scanning area is 950 mm by 950 mm. This adds up to 32000 individual images, or in total, up to 700 GB of disk space.

To reduce the amount of data, images can be deleted after they are analyzed. This can be done while the scanning is underway. In this way, data reduction of roughly a factor of 40 is achieved. In addition, small thumbnail images of objects belonging to a certain class can be optionally recorded on disk for later inspection.

To take full advantage of the data reduction, the analysis has to be fast enough to keep up with the scanning speed. The time to analyze a single image depends on the complexity of the preprocessing and the segmentation algorithms. The analysis can be done in parallel, so that several images in different phases are computed at the same time on different cores of a multicore processor. The memory used by the collected data is also managed with a disk caching system to reduce the memory usage needed for the large GEM foils. This is especially important on computers with 32-bit architecture, where the amount of reservable memory is limited.

\section{Test scans}

\subsection{Setup}

\begin{figure}\centering
\includegraphics[width=0.48\textwidth]{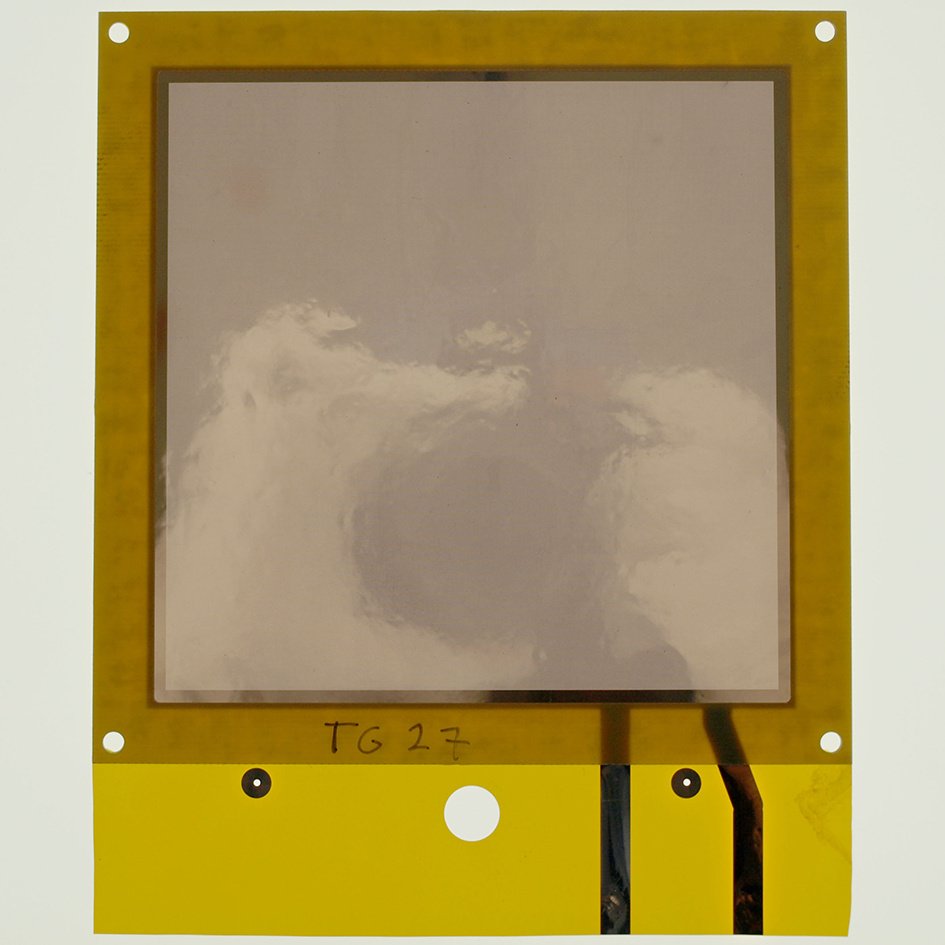}
\caption{The Top side of the foil. The identification of the orientation is made based on the location of the two strips seen on the lower right corner of the picture.\label{foilorientation}}
\end{figure}

A set of five 10 cm by 10 cm GEM foils was scanned from both sides with the scanner system. The foils were CERN standard double mask GEM foils with pitch of 140~$\mu$m and outer- and inner hole diameters of 70~$\mu$m and 50~$\mu$m respectively. The foils were analyzed to measure the uniformity of the foil characteristics. The foils were oriented on the scanning table as seen in Figure~\ref{foilorientation}. The face seen in Figure~\ref{foilorientation} was labeled Top side. Each foil was turned around from left to right to scan the Bottom side.

\begin{figure}\centering
\includegraphics[width=0.48\textwidth]{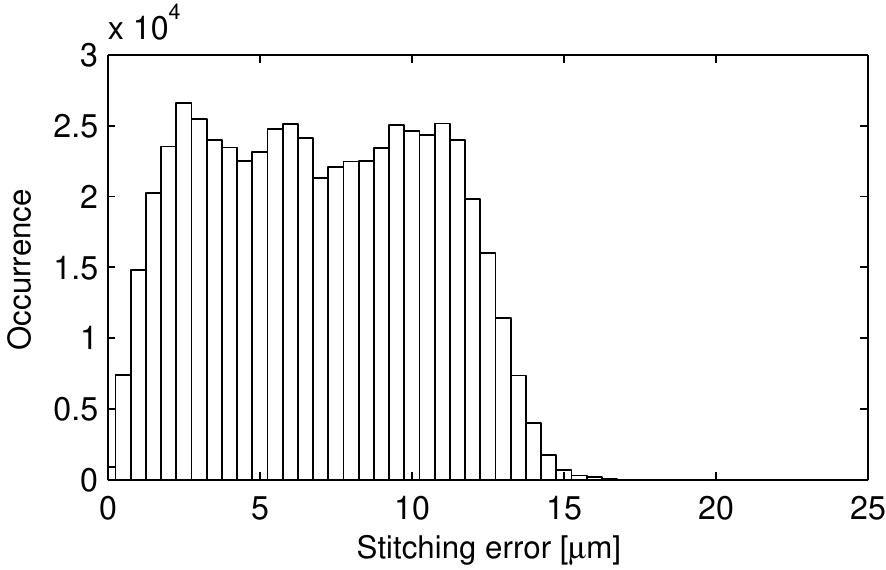}
\caption{Distance between the centroids of corresponding holes from the Top and Bottom sides.\label{stitchingerror}}
\end{figure}

Small random rotations between the images were found. This was most likely due to small rotations in the movement of the camera along the z-axis during the image focusing process. The rotation caused a random shift of a few pixels between neighboring images which in turn introduced an error on the stitching. Due to this error, the centroids of the holes, measured from both sides, did not exactly match each other. A histogram of the magnitude of this shift in the centroids, measured from the inner holes, is shown in Figure~\ref{stitchingerror}. For the 10 cm by 10 cm foils, the shift was less than the radius of a typical inner hole.

The measurement of the GEM foil characteristics was done by using the ellipse fit parameters calculated by the analysis software. These are the ellipse major and minor axes, denoted as a and b, respectively, and the angle between the ellipse major axis and the x-axis of the image. The diameter of a hole was calculated as the average value of a and b. The difference between a and b, or axial difference, was used to describe the degree of ellipticity of the features.

\subsection{Measurement uncertainty}

The measurement uncertainty of the system was estimated by using the data from the measurement of the inner holes, since this should be identical between the sides of the foils. Each hole was compared with the corresponding measurement from the other side. The data was combined by selecting three corner holes from both the Top and the Bottom data sets. Both data sets were rotated to level them precisely. Subsequently, one of the data sets was reflected along the y-axis and compared with the other. Holes that fell within 40~$\mu$m of each other were identified as corresponding to the same physical hole. Holes on one side without a corresponding hole on the other side were discarded from the analysis.

\begin{figure}\centering
\includegraphics[width=0.48\textwidth]{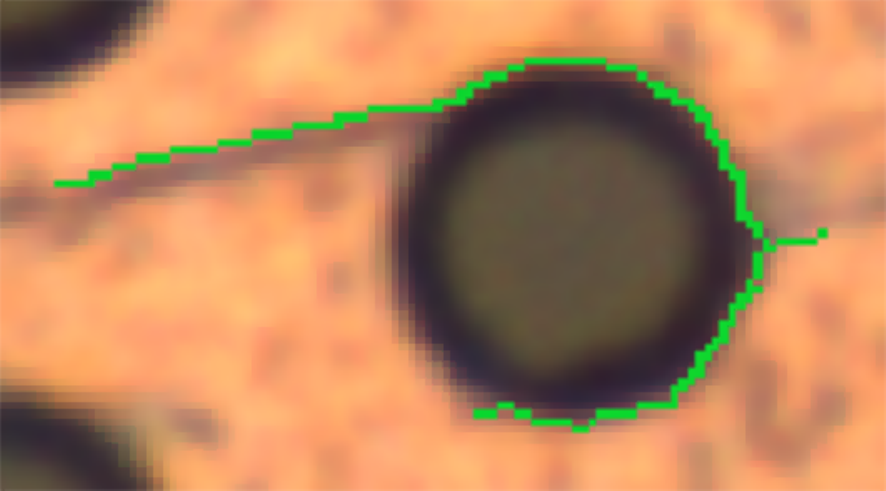}
\caption{A badly defined contour of an outer hole.\label{contour}}
\end{figure}

Object classification by neural network introduces occasional errors in the classification giving either a false positive or a false negative result. However, the demand of a corresponding hole on both sides effectively filters out incorrectly classified features from the analysis. The surface texture of the foil occasionally interferes with the detection of outer holes as sometimes the edge detector starts to follow a high contrast feature on the copper surface. Furthermore the edge detector can not always follow the outer hole fully because of highly contrasting features near the boundary resulting in a badly defined contour. An example of a bad contour is shown in Figure~\ref{contour}. These incorrectly defined outer holes have been excluded from the analysis by the classification algorithm.  Altogether 99.7~\% of inner holes and 76.4~\% of outer holes were found and classified correctly.

\begin{figure}\centering
\includegraphics[width=0.48\textwidth]{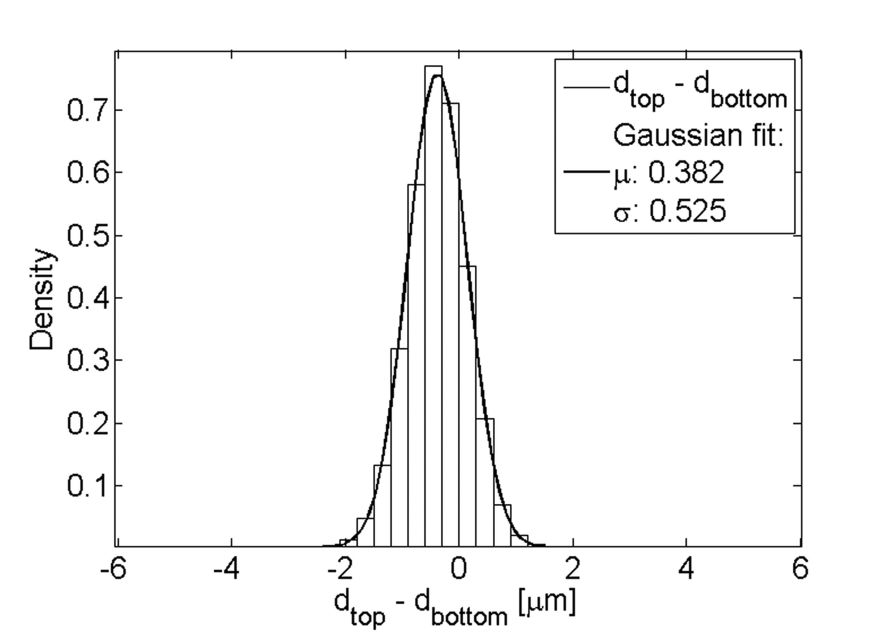}
\caption{Difference between the inner hole diameters of corresponding holes measured from the Top and the Bottom sides.\label{innerdiameteruncertainty}}
\end{figure}

The uncertainty of the diameter, the axial difference and the angle were evaluated from the inner hole data. The angle data of the Bottom data set was subtracted from 180$^\circ$ to account for the turning over of the foil. The distribution of the Top diameter minus the Bottom diameter of each inner hole for foil number 1 is shown in Figure~\ref{innerdiameteruncertainty}. The mean value of this distribution is expected to be zero, but it was found to vary, with a standard deviation of 0.34~$\mu$m (calculated from the similar distributions of all the 5 foils). This uncertainty is probably due to minute differences in the lighting of the images which affects the edge detection algorithm when measuring the thin transparent polyimide inner boundary. This was understood as a systematic uncertainty of the diameter measurement. The outer holes could not be measured in a similar way, but such uncertainty was not found in the outer hole data when comparing two scans from the same side of a foil.

The pitch, the diameter and the axial difference measurement uncertainty were calculated from the standard deviation of the difference between the inner hole data of the Top and the Bottom side measurements. The pitch measurement value includes error of the stitching in image boundaries. Total uncertainties, including systematic uncertainty, for the measurements was then found to be 0.3~$\mu$m, 0.5~$\mu$m and 0.6~$\mu$m for pitch, diameter and axial difference, respectively. The uncertainty for relative diameter differences within a single scan is 0.4~$\mu$m, however.

The mean axial difference of the holes in all the foils except in the foil 3 was between 1.60~$\mu$m and 2.17~$\mu$m, or on the order of one pixel (1.75~$\mu$m). This was understood as a bias in the ellipse fitting algorithm, which had a tendency to find elliptical shapes even when the holes were nearly perfect circles.

\begin{figure}\centering
\includegraphics[width=0.48\textwidth]{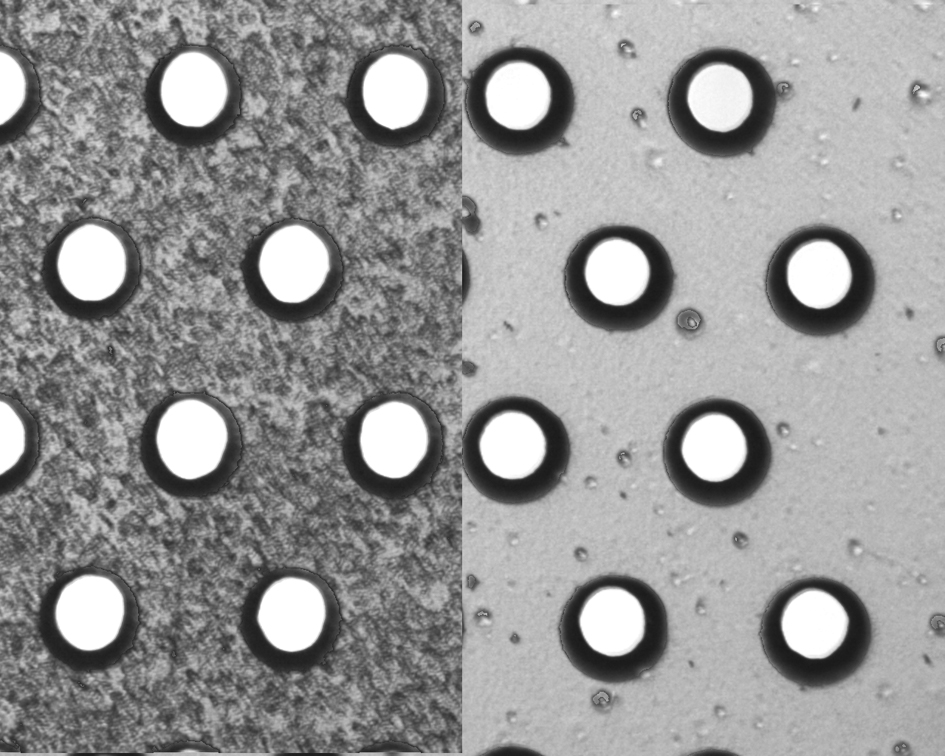}
\caption{A comparison of microscope images of the holes in foil 3 (left) and foil 5 (right).\label{microscopecomparison}}
\end{figure}

\begin{figure}\centering
\includegraphics[width=0.48\textwidth]{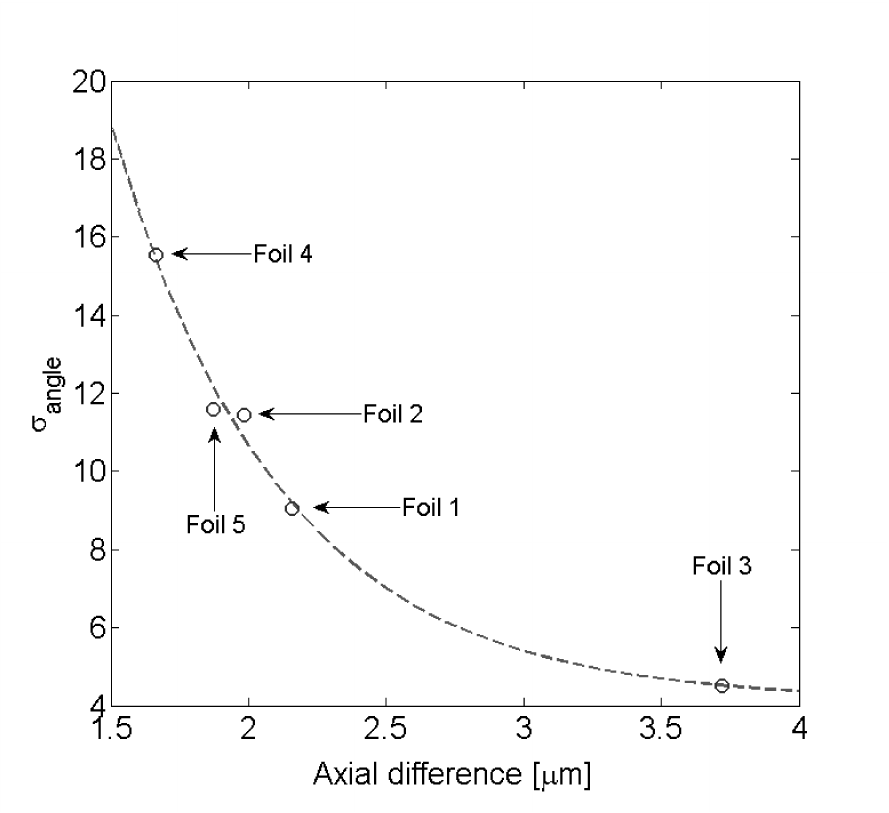}
\caption{The uncertainty of the angle measurement vs. the average axial difference of the holes for all foils. An exponential curve has been added to guide the eye.
\label{angleuncertainty}}
\end{figure}

The axial difference and the uncertainty of the measured angle were correlated so that the angle measurement was more accurate when the mean axial difference was large. This makes sense as the angle of the ellipse would become arbitrary in the limit of the hole being circular. When the holes were visibly elliptical, as in the foil 3 on the left side of Figure~\ref{microscopecomparison}, the mean axial difference was 3.72~$\mu$m and the angle measurement uncertainty diminished by a factor of three, as can be seen from Figure~\ref{angleuncertainty}. Thus, the axial difference measurement can be used as an indication of elliptical holes when the value significantly exceeds a single pixel width.

\subsection{Results}

\begin{figure*}\centering
\includegraphics[width=0.9\textwidth]{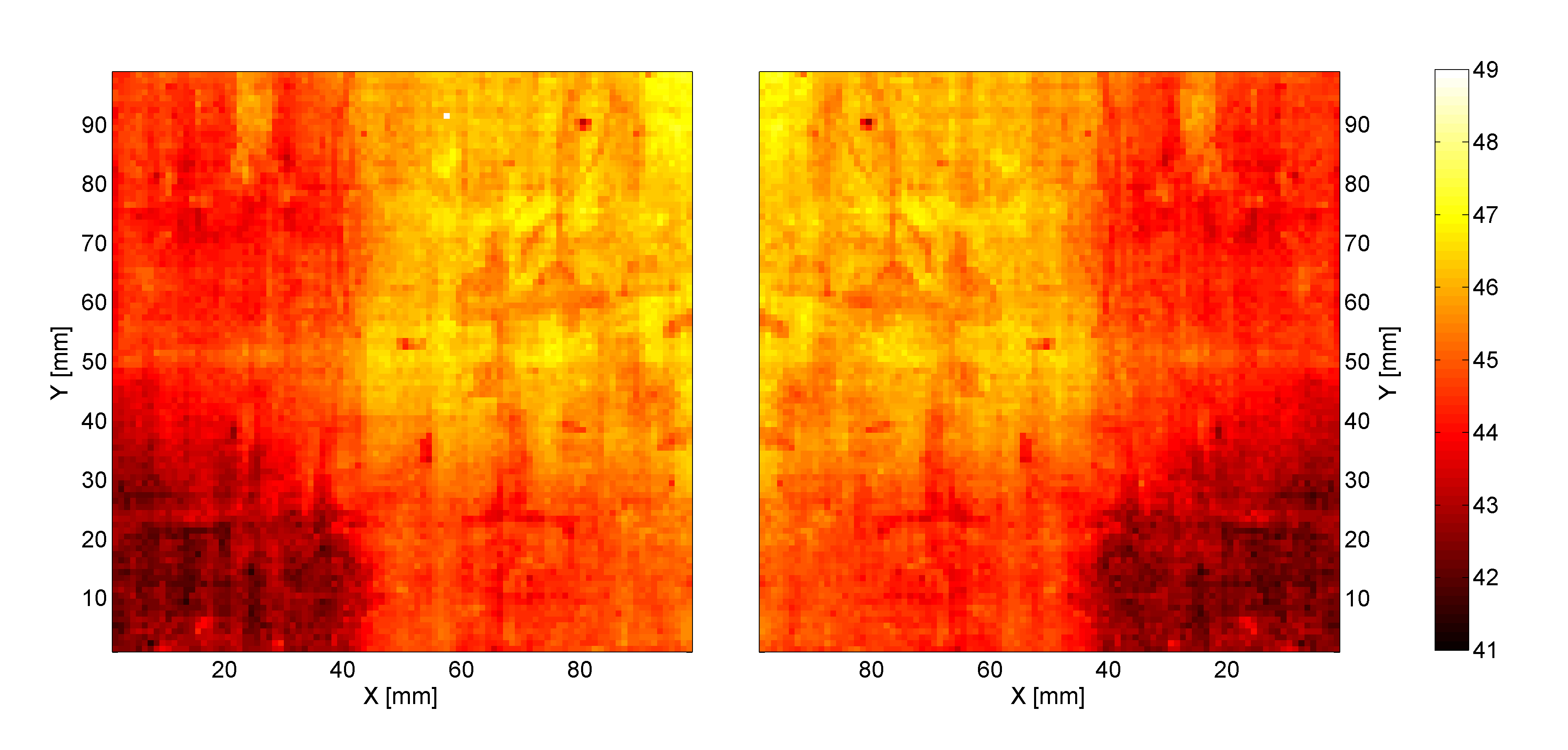}
\caption{Inner hole diameter histograms from Top (left) and Bottom (right) sides of foil 5. Bin width is 1 mm$^2$. Diameters are in $\mu$m.\label{innerholefoil5tb}}
\end{figure*}

\begin{figure*}\centering
\includegraphics[width=0.9\textwidth]{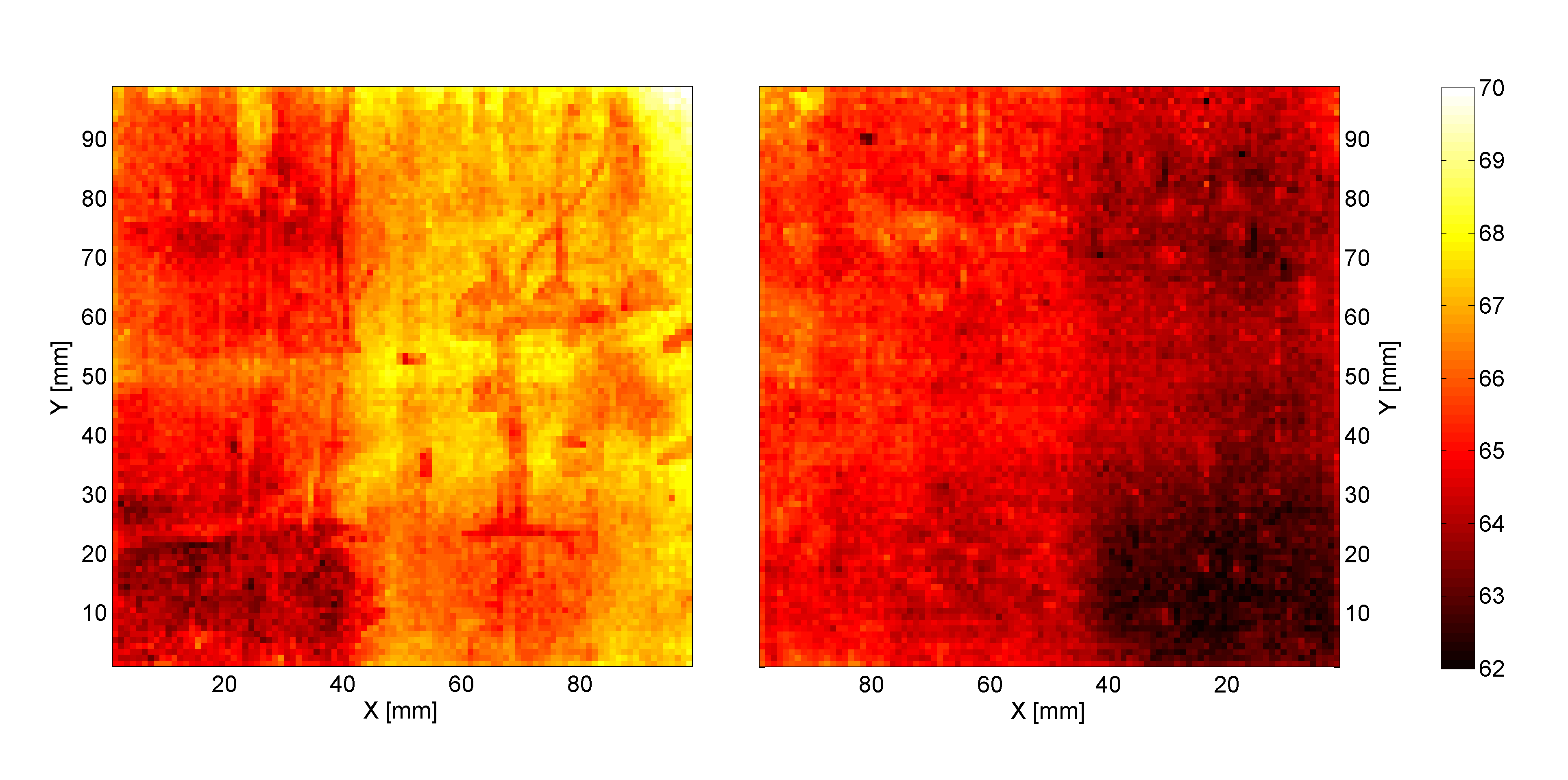}
\caption{Outer hole diameter histograms from Top (left) and Bottom (right) sides of foil 5. Bin width is 1 mm$^2$. Diameters are in $\mu$m.\label{outerholefoil5tb}}
\end{figure*}

Example histograms of the inner and the outer holes from both sides of one foil are shown in Figures~\ref{innerholefoil5tb} and ~\ref{outerholefoil5tb}. 

\begin{figure}\centering
\hspace{-1.5em}
\includegraphics[width=.45\textwidth]{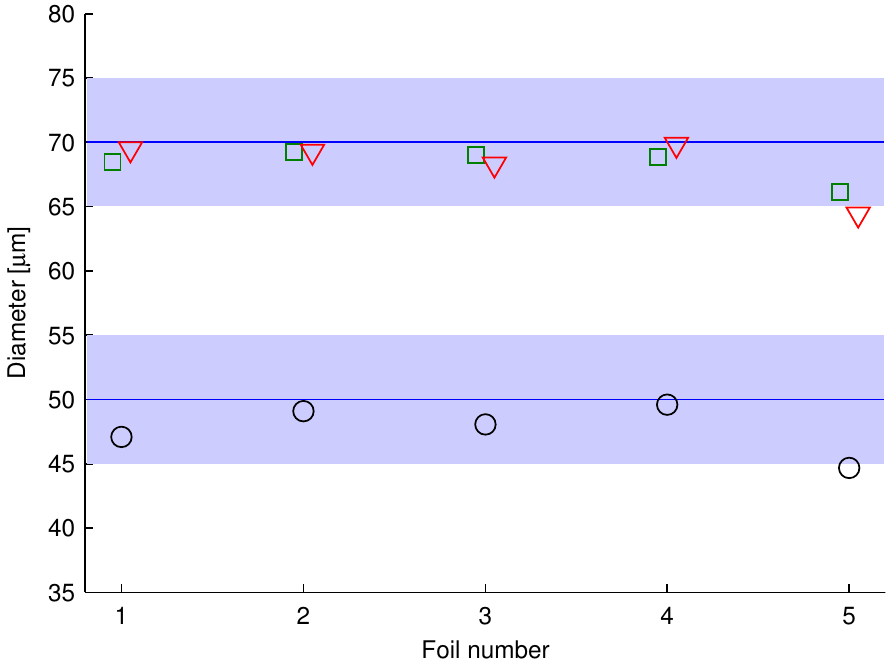}
\caption{The mean diameter of the inner holes (circles) and the outer holes (squares for top, triangles for bottom) of each foil. Shaded area represents the tolerance given by manufacturer of the foils.\label{meandiameter}}
\end{figure}

It can be seen that the features that are visible in the outer hole measurement are reflected in the diameters of inner holes. The inner hole histograms are practically identical. The mean diameters of the holes, both the inner and the outer, for each foil are shown in Figure~\ref{meandiameter}.

\begin{figure}\centering
\includegraphics[width=0.45\textwidth]{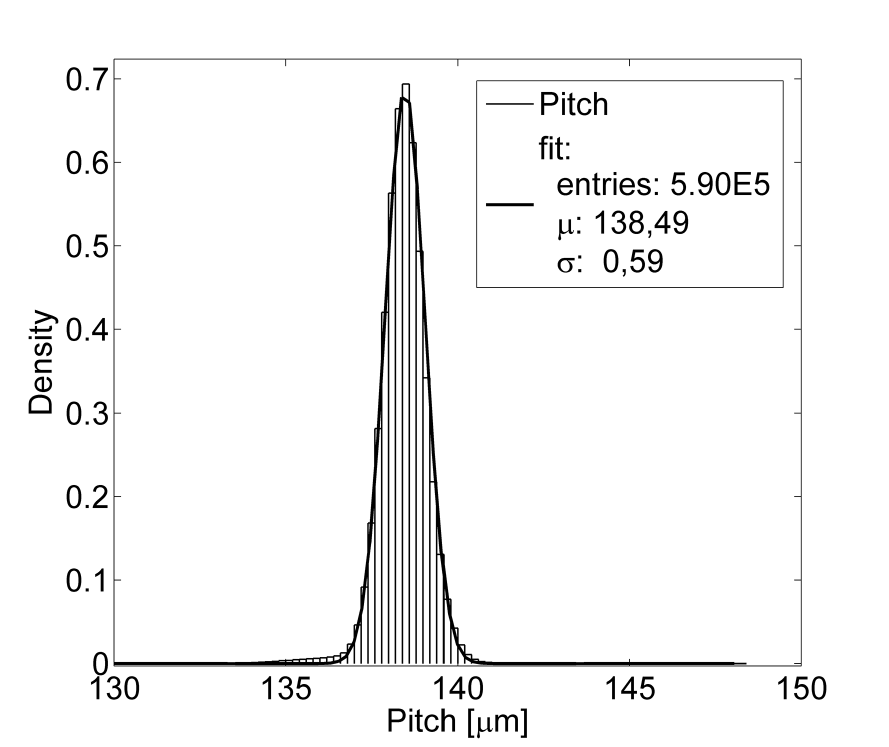}
\caption{Histogram of the pitch of foil 1.\label{pitchfoil1}}
\end{figure}

The pitch of the foils was measured by Delaunay triangulation of the centroid coordinates of the inner holes. A histogram of the pitch, or the lengths of the sides of the triangles, of the foil 1 is shown in Figure~\ref{pitchfoil1}. The pitch was uniform over all the foils with a mean of 138.5~$\mu$m and a standard deviation of 0.6~$\mu$m.

\section{Comparison with gain measurements}

A gain mapping was done for one of the foils, foil 2, to correlate the hole size variation with the measured variation in the gas multiplication. The aim of the exercise was to emphasize the motivation for the hole size measurement. Two different measurement setups were used. A high precision setup (setup 1) was utilized with a low gain over the foil and a lower precision setup (setup 2) with high gain and different gas mixture. The latter setup was used to see if any correlation would hold with a radically different setup.

\subsection{Setup 1}

\begin{figure*}\centering
\subfigure[]{
\includegraphics[width=0.44\textwidth]{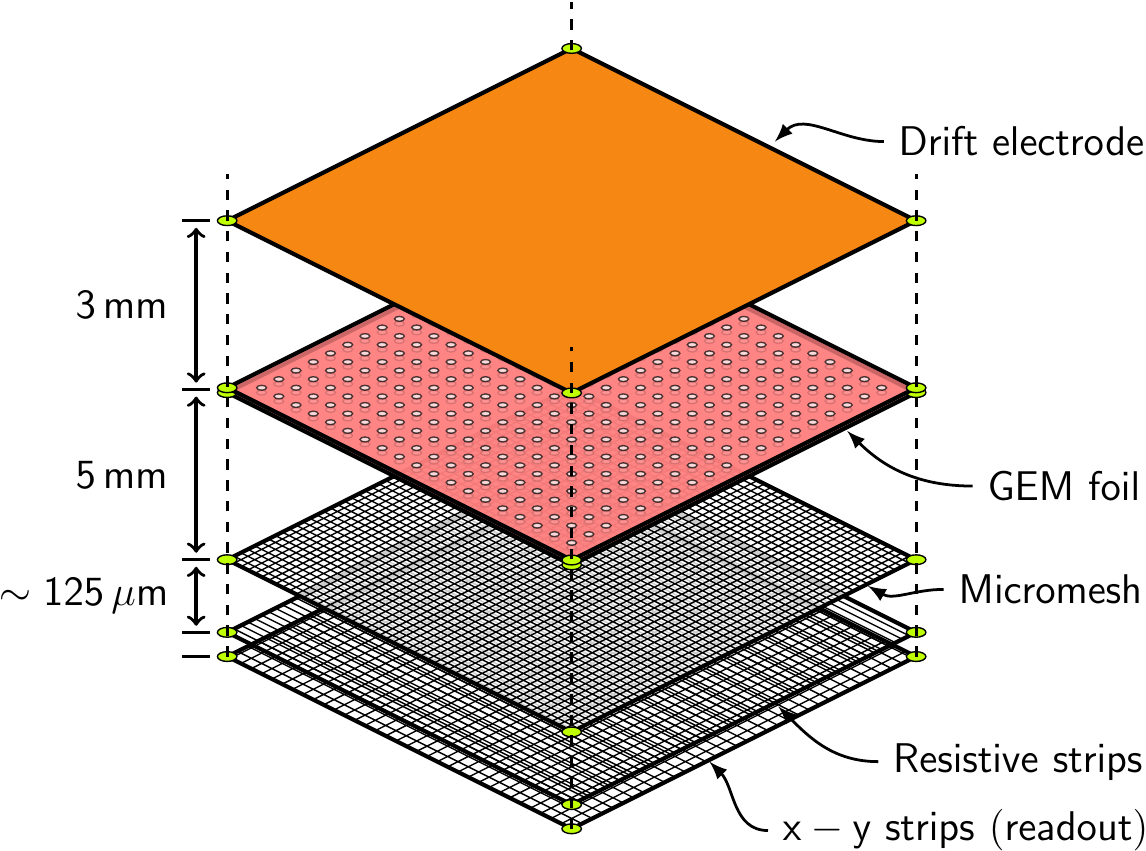}
}
\subfigure[]{
\includegraphics[width=0.42\textwidth]{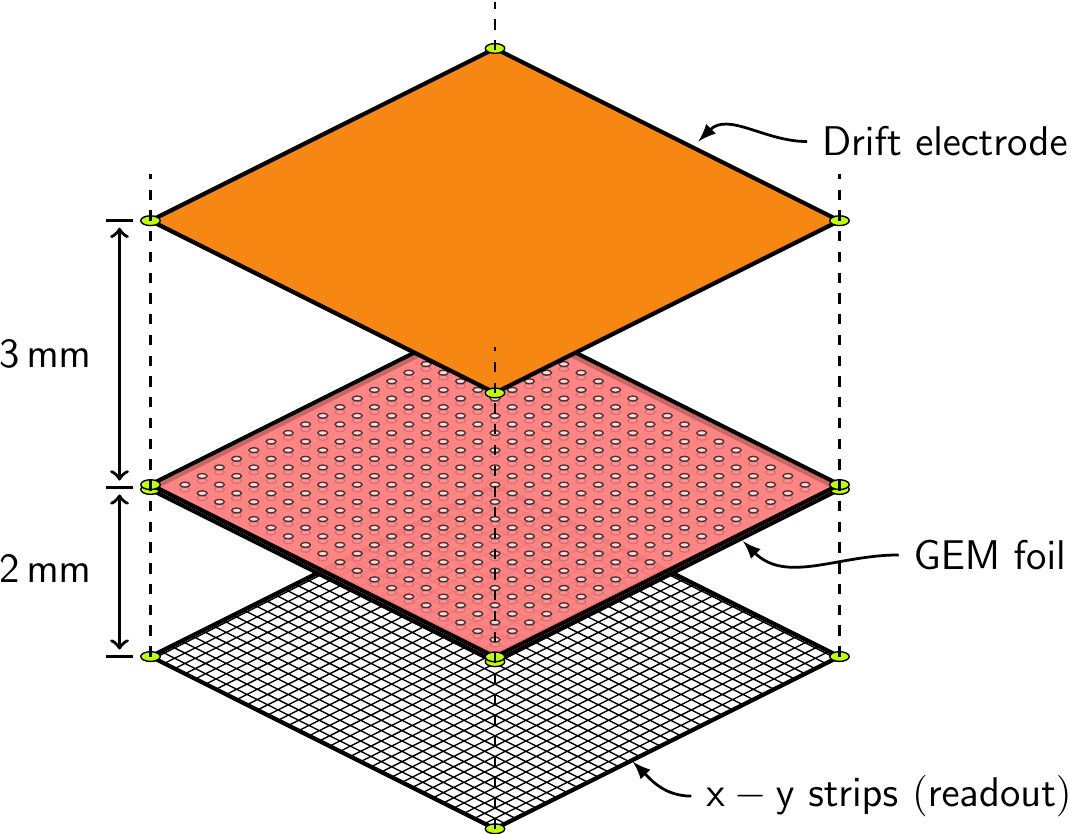}
}
\caption{Schematic view of the gain measurement setups. (a) Setup 1 using a MICROMEGAS detector. (b) Setup 2 using single GEM foil,\label{mmggemsetup}}
\end{figure*}

In setup 1, the GEM foil was assembled on top of a Micro-Mesh Gaseous Structure (MICROMEGAS) detector~\citep{Giomataris:1995fq}. The transfer gap between the mesh and the GEM foil was 5 mm and the drift gap between the foil and the drift electrode was 3 mm.  A mixture of 90~\% argon and 10~\% CO$_2$ was used as measurement gas. The GEM voltage was supplied through a chain of resistors. The Voltage over the GEM foil was set to 267~V. The drift field was 555~V/cm and the transfer field between the mesh and the bottom electrode of the GEM foil was 534~V/cm. A positive 500~V high voltage was supplied to the resistive strips to power up the MICROMEGAS detector. A schematic view of the setup is shown in Figure~\ref{mmggemsetup}.

The GEM foil was operated approximately at a gain of 2 and was irradiated with a $^{55}$Fe source. Two peaks could be seen due to the low gain of the GEM foil. The lower peak corresponds to primary ionization in the transfer region between the MICROMEGAS mesh and the GEM foil and the higher to that in the drift region of the detector. Any local variation in the performance of the underlying MICROMEGAS detector was thus present in both peaks and could be eliminated by inspecting the ratio of the two peaks.

The MICROMEGAS detector used in this setup had a 360 by 360 strip resistive x-y readout over a 90 mm by 90 mm area. The detector was read out with 6 APV chips, reading all the 360 channels from both the x- and y- strips. A preamplifier was used to read the signal on the MICROMEGAS mesh. The mesh signal was used for triggering the system. A $^{55}$Fe source was positioned above the detector and left there for several days while the detector was powered on and constantly irradiated. Approximately \mbox{$2\times 10^6$}~events were collected in total over several measurement runs during the time.

The mesh of the MICROMEGAS detector is supported by a grid of supporting pillars, which causes a periodic pattern of dead area within the active area of the detector. The data was analyzed in a grid of 20 by 20 bins (4.5 mm by 4.5 mm). The bin size was larger than the spacing of the pillars, so possible effects due to the spacers were not considered. 

\begin{figure}\centering
\includegraphics[width=0.46\textwidth]{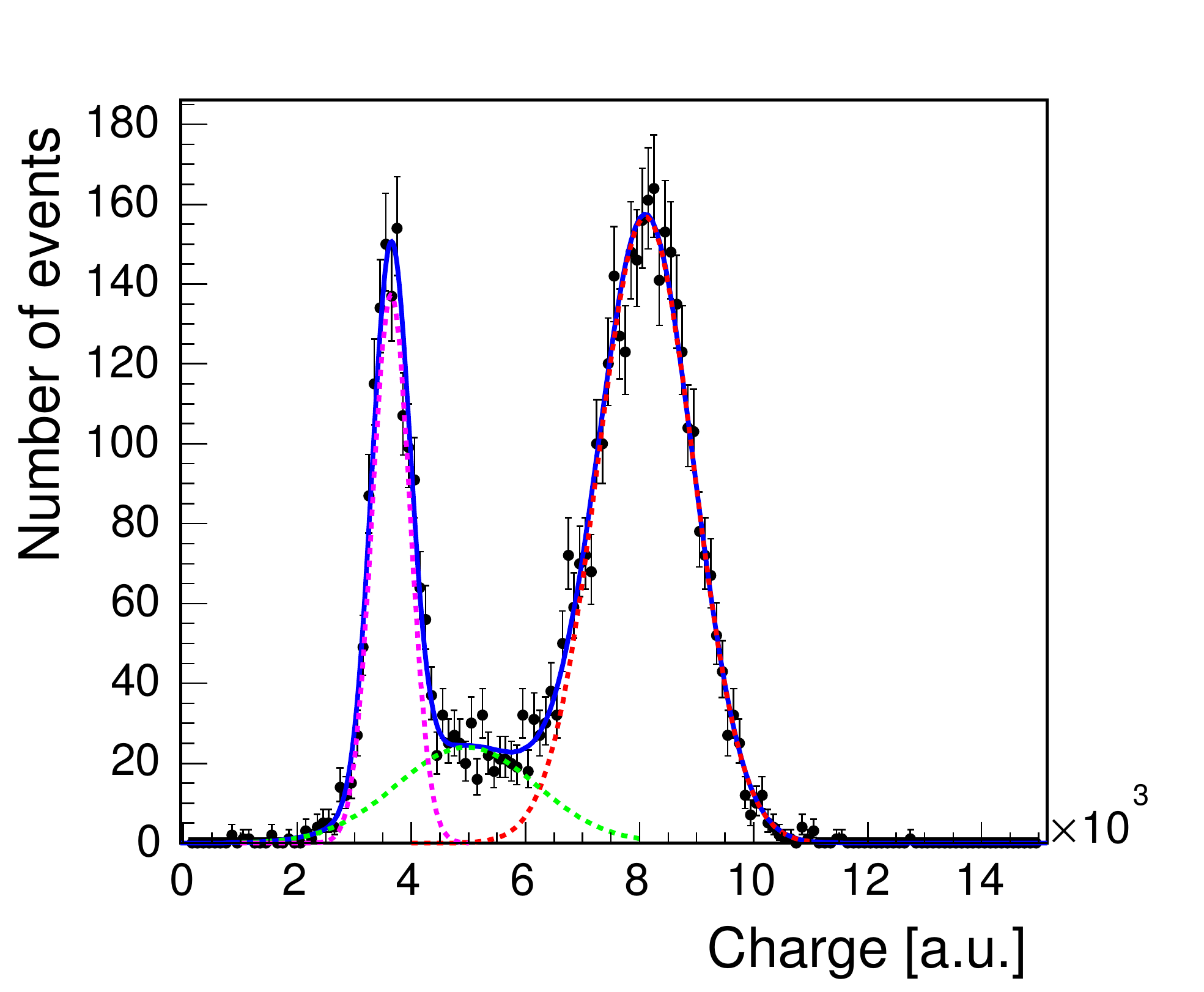}
\caption{Charge distribution for a typical detector bin. The data was fitted using three Gaussian functions. The peak in the middle accounts for the GEM escape peak. \label{tg24charge}}
\end{figure}

\begin{figure}\centering
\includegraphics[width=0.46\textwidth]{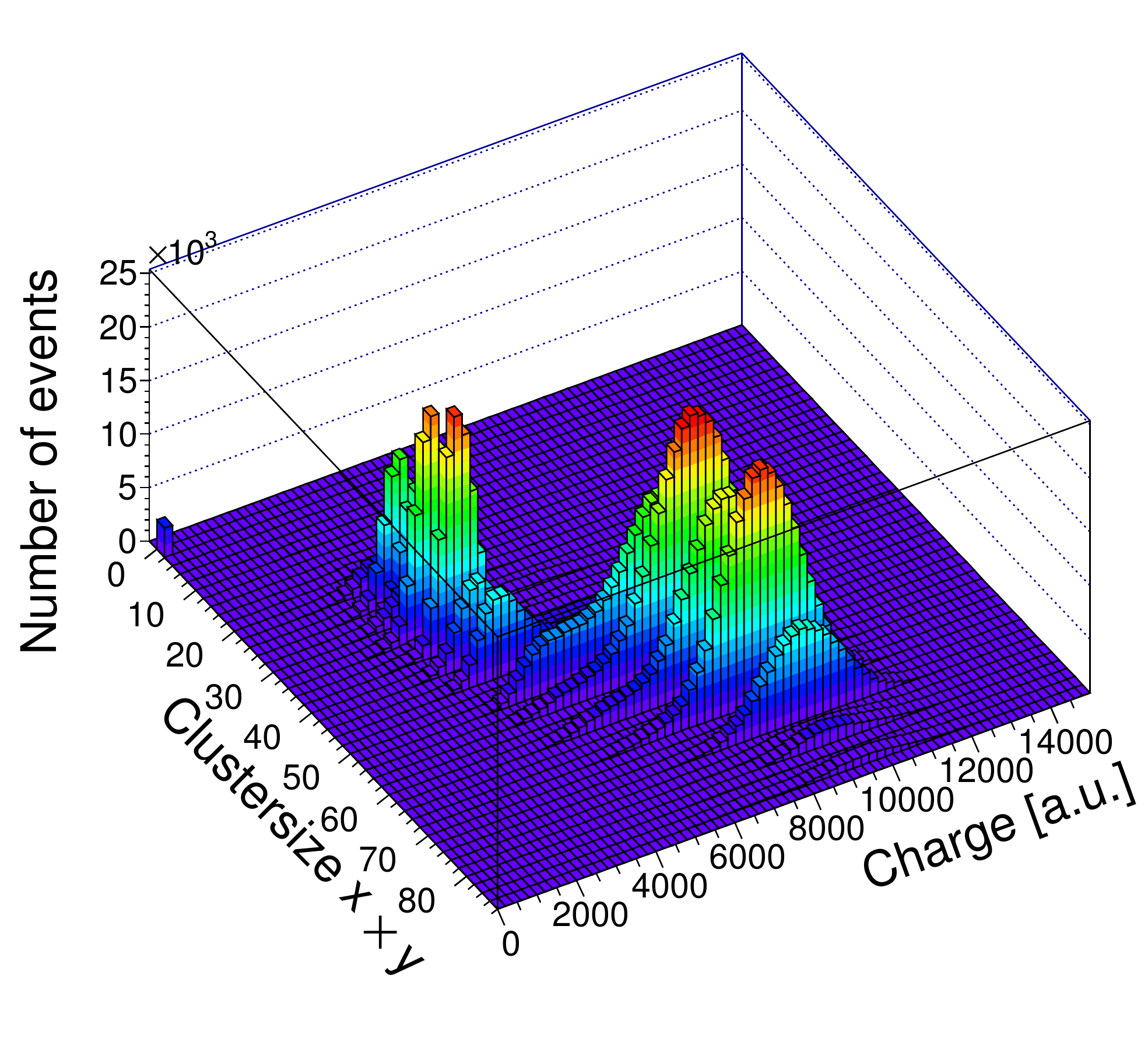}
\caption{Charge vs. cluster area size for the full detector.\label{tg24cluster}}
\end{figure}

An example histogram of the collected charge of one bin is shown in Figure~\ref{tg24charge}. Gaussian functions were fitted separately to the GEM peak and the MICROMEGAS peak. The gain of the GEM foil was calculated by determining the ratio of the centroids of the two $^{55}$Fe peaks. The escape peak of the GEM signal overlaps with the main peak of the MICROMEGAS signal. This was unavoidable due to the narrow dynamic range achieved with the APV chips. The GEM signal was, however, spread over a larger area due to diffusion. The cluster size of the GEM events was larger as can be seen in Figure~\ref{tg24cluster}. A portion of the escape peak could be cut from the data by a combined charge and cluster area cut. 
To estimate the error caused by the overlapping GEM escape peak, charge distributions without any cuts of ten random bins were fitted by three Gaussian functions to account for the escape peak. The edge bins were excluded. The uncertainty was found to be 0.015, roughly 1~\% of the mean value of the gain.

\subsection{Setup 2}

Foils 1, 2 and 5 were later measured with a second setup. The foils were used in a single-foil GEM detector and the spectrum from a collimated $^{55}$Fe source was measured. A mixture of 70~\% argon and 30~\% CO$_2$ was used as measurement gas. In the second setup the induction field was set to 2600~V/cm and the drift field was 470~V/cm. The voltage over the GEM foil was 526~V, reaching a gain of approximately 600. The signal was read from the top strips of the x-y strip readout. All 128 strips of a connector were connected to a preamplifier as a single channel. 

A sheet of plastic, 3 mm thick, with a regular 8 by 8 hole grid of holes each 2 mm in diameter was used as a collimator. The pitch of the holes was 12.5 mm. The preamplifier was switched between connectors as the scan proceeded to the corresponding half of the detector. All the other strips were kept terminated throughout the measurement. 

In the second setup the foil could not be kept in constant irradiation throughout the measurement. This fact combined with the radically higher gain of the foil lead into observable charge-up behavior in the foil. During charge-up of a fresh foil the measured gain increased by approximately \mbox{10~\% - 15~\%}. Each measurement point was irradiated with an $^{241}$Am source with a rate of approximately 1,6 kHz for 30 seconds prior to the measurement with $^{55}$Fe. This was done to introduce a significant amount of ionization into the detector volume prior to the measurement to reduce charge-up related variation in the measurements. 

Due to challenges to control the placement of the source on the collimator plate the position where the collimated beam of x-rays hit the foil in each measurement point was not perfectly precise. Instead, the foil data was divided into 64 12.5 mm by 12.5 mm bins, each corresponding to a hole in the collimator plate. The gain of the edge bins varied significantly under repeated measurements of the same foil. This was assumed to be due to imprecise source placement and edge effects of the detector. The edge bins were excluded from the analysis, leaving six measurements of the 6 by 6 central bins, or 216 single measurements.

\subsection{Results}

\begin{figure}[htp]\centering
\includegraphics[width=0.45\textwidth]{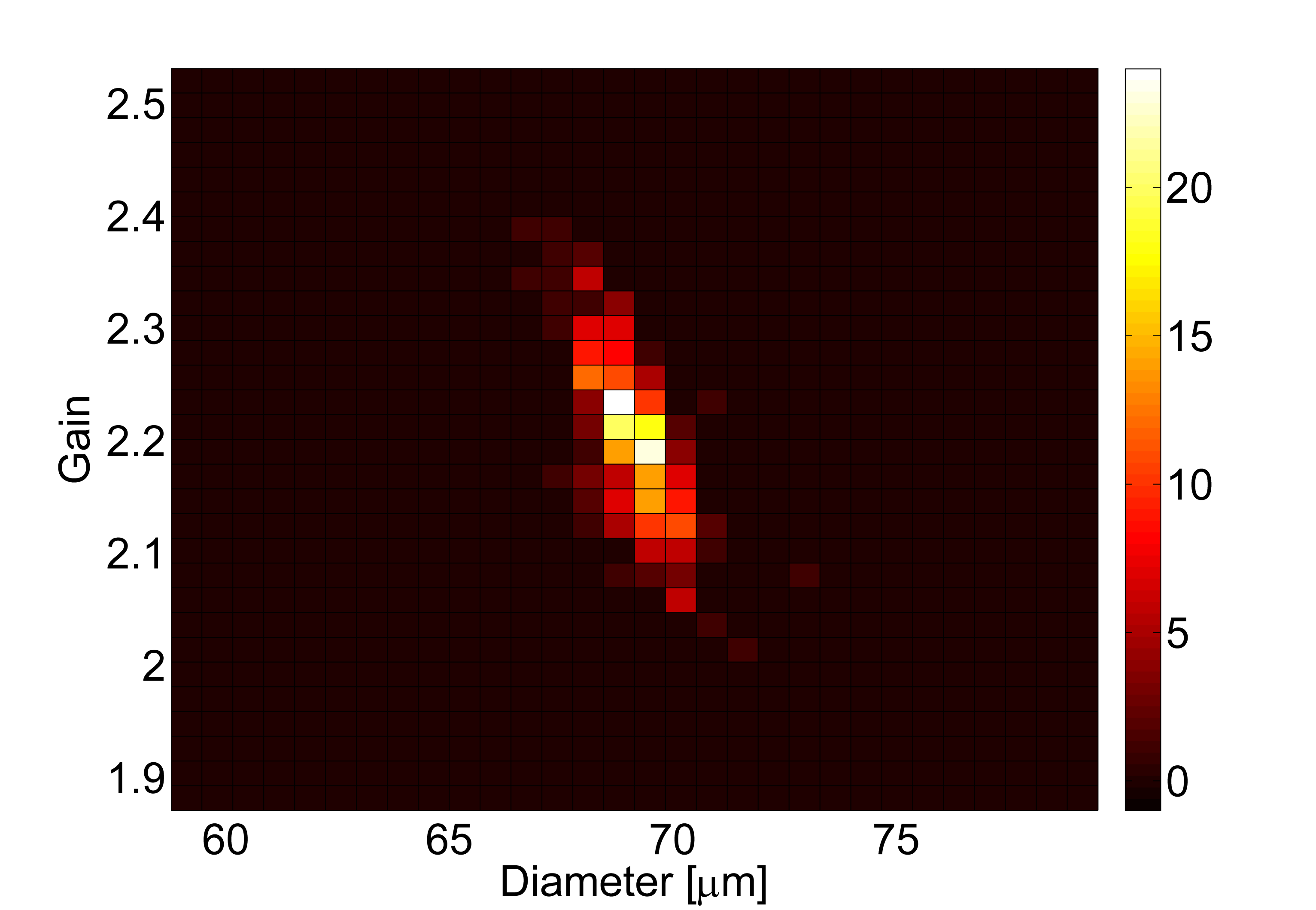}
\caption{A histogram of the correlation between the measured gain and the outer top hole diameters. The gain was measured with setup 1.\label{outertop_correlation}}
\end{figure}

A histogram of the correlation of the measured outer hole diameters of the top side of the foil against the measured gain of foil 2 is shown in Figure~\ref{outertop_correlation}. The gain was measured with setup 1. The plot suggests an inverse dependence of gain on the diameter. The inner holes and outer holes of the bottom side show a similar correlation in their behavior. 

\begin{figure}[htp]\centering
\includegraphics[width=0.45\textwidth]{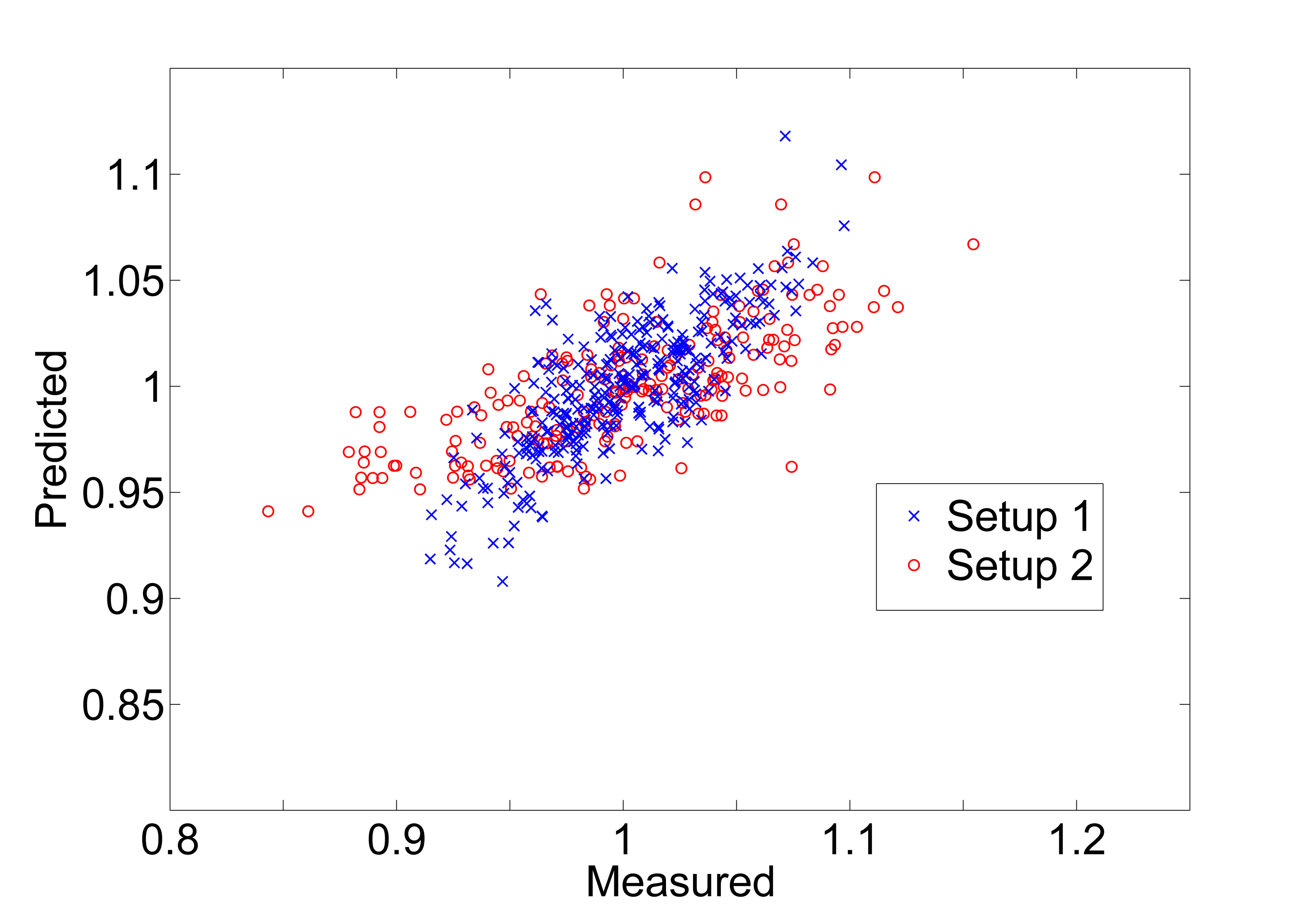}
\caption{The gain variation prediction based on inverse linear correlation on diameter plotted against the measured relative gain. The measurement with setup1  is drawn with crosses and the measurement with setup 2 with circles.\label{dumb_correlation}}
\end{figure}

A simplified prediction of the gain variation was made by assuming an inverse linear dependence on the hole diameters. The relative gain in each bin was predicted by normalizing the median values of the inner and outer hole diameters each by the mean of all of the bins and multiplying the inverse values together. A correlation plot of the prediction against the relative gain of measurements made with both setups is shown in Figure~
\ref{dumb_correlation}.

It can be readily seen in Figure~\ref{dumb_correlation} that the second set of measurements behaves differently due to the fact that the foils were operated in a region of significantly higher gain and with different gas mixture. Additionally, the size variation and thus the prediction is flatter because of the large bins. A more accurate prediction of the gain would have to take into account the exponential increase of the gain of the GEM foil as a function of the voltage. 

\begin{figure*}[htp]\centering
\subfigure[]{
\includegraphics[width=0.47\textwidth]{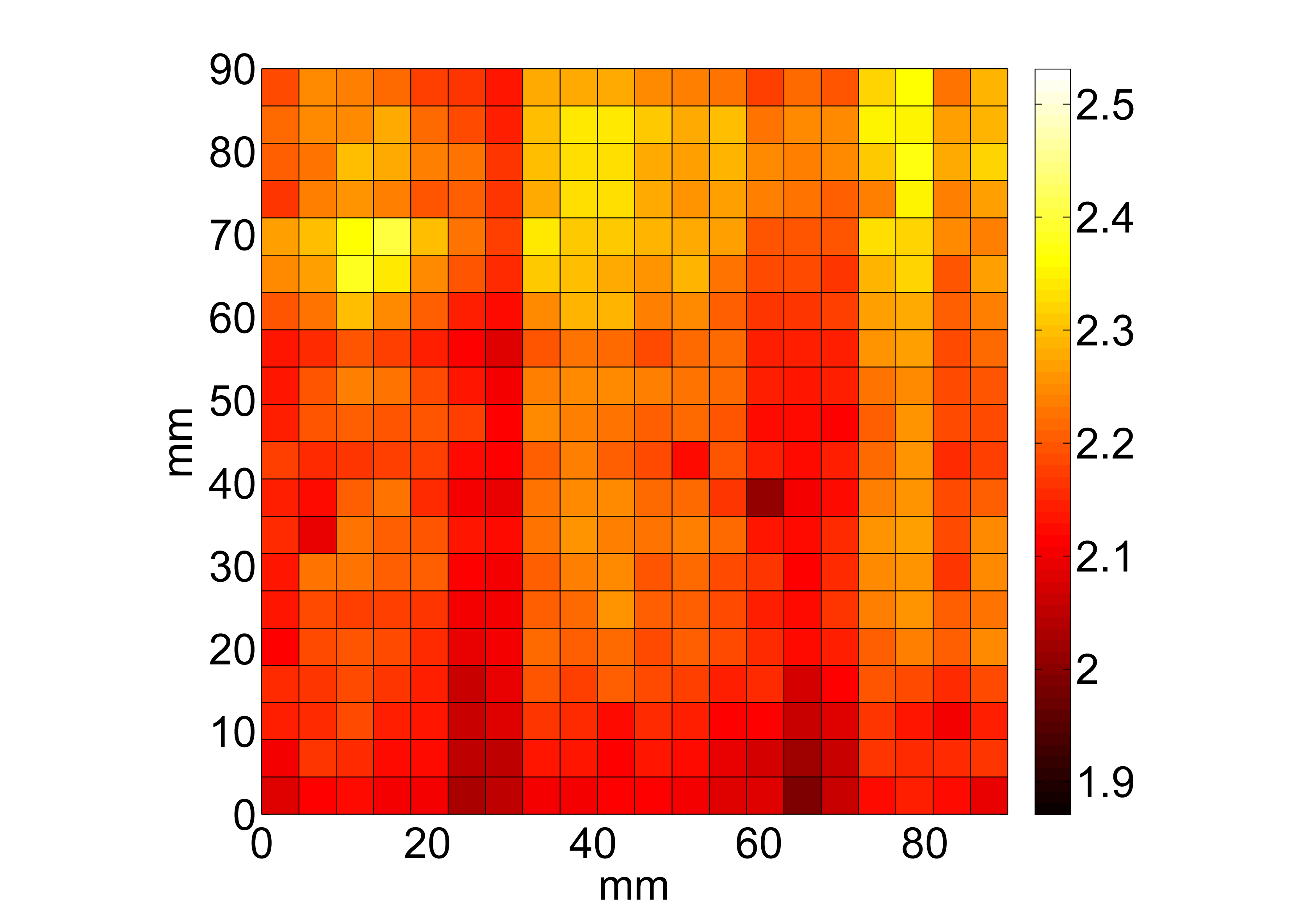}
}
\subfigure[]{
\includegraphics[width=0.47\textwidth]{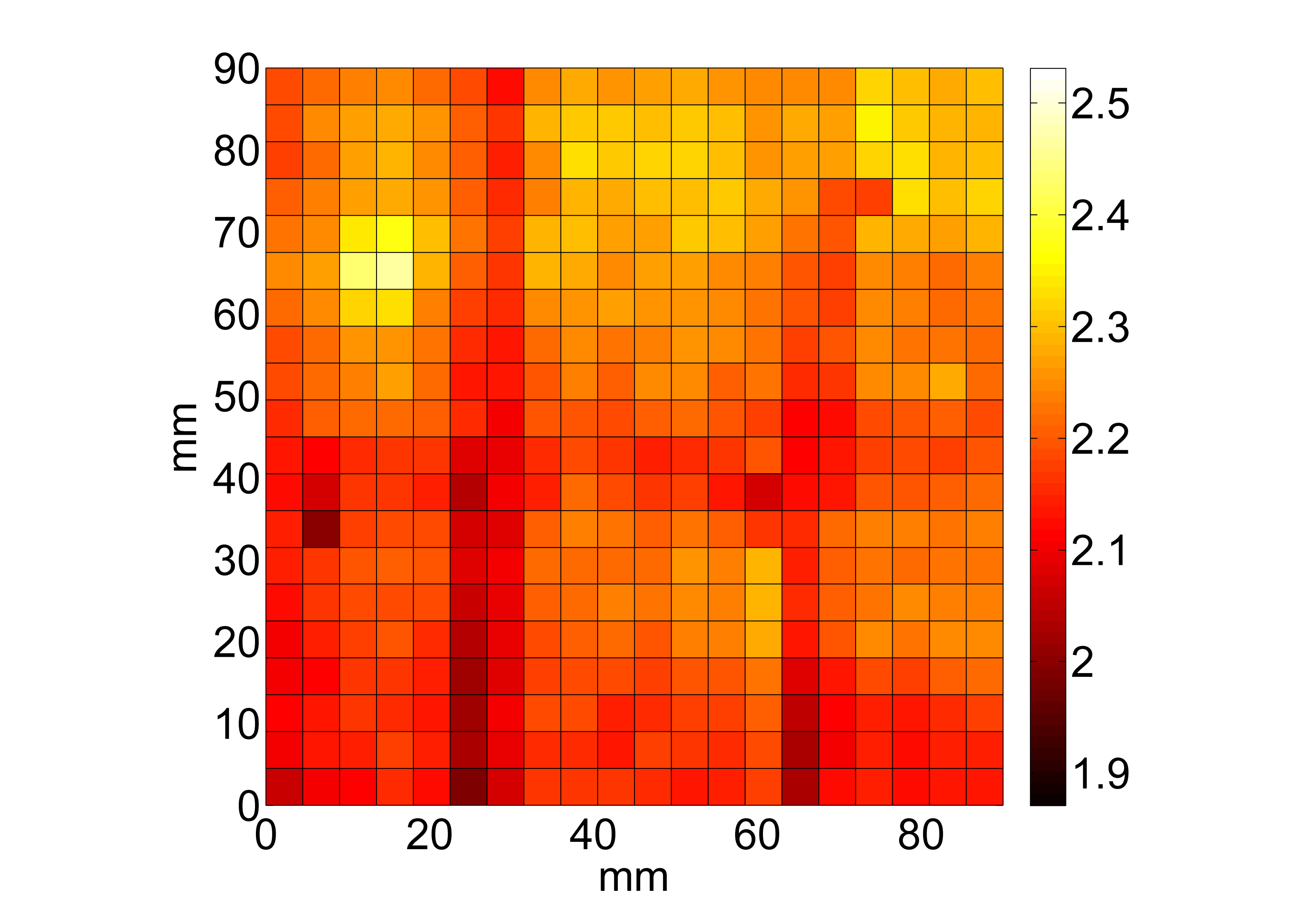}
}
\caption{2d plot of the gain measured with setup 1 (a) and the predicted gain (b) of foil 2.\label{measured}}

\end{figure*}

The gain measured with setup 1 and the prediction are shown in Figure~\ref{measured}(a) and (b). Even though the prediction does not give the absolute gain of the GEM foil, it can be used to estimate the relative gain variation. The prediction, based on hole size measurement, can thus be used in QA to avoid the stacking up of foils with overlapping areas of extreme inhomogeneities in multiple foil GEM detectors.

\begin{figure}[htp]\centering
\includegraphics[width=0.47\textwidth]{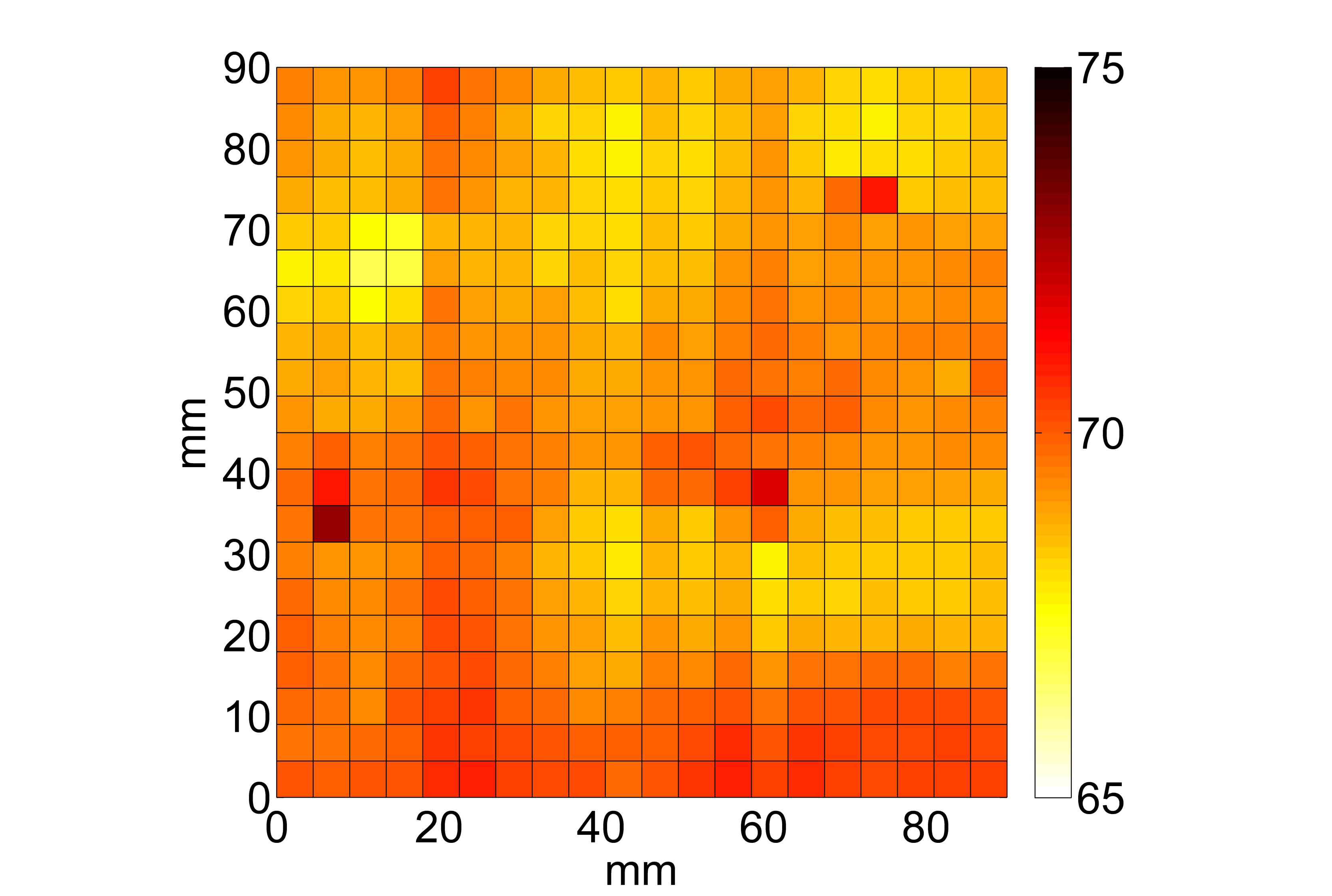}
\caption{2d histogram of the measured diameters of the outer holes on the top side of foil 2.\label{outertop}}
\end{figure}

\begin{figure}[htp]\centering
\includegraphics[width=0.47\textwidth]{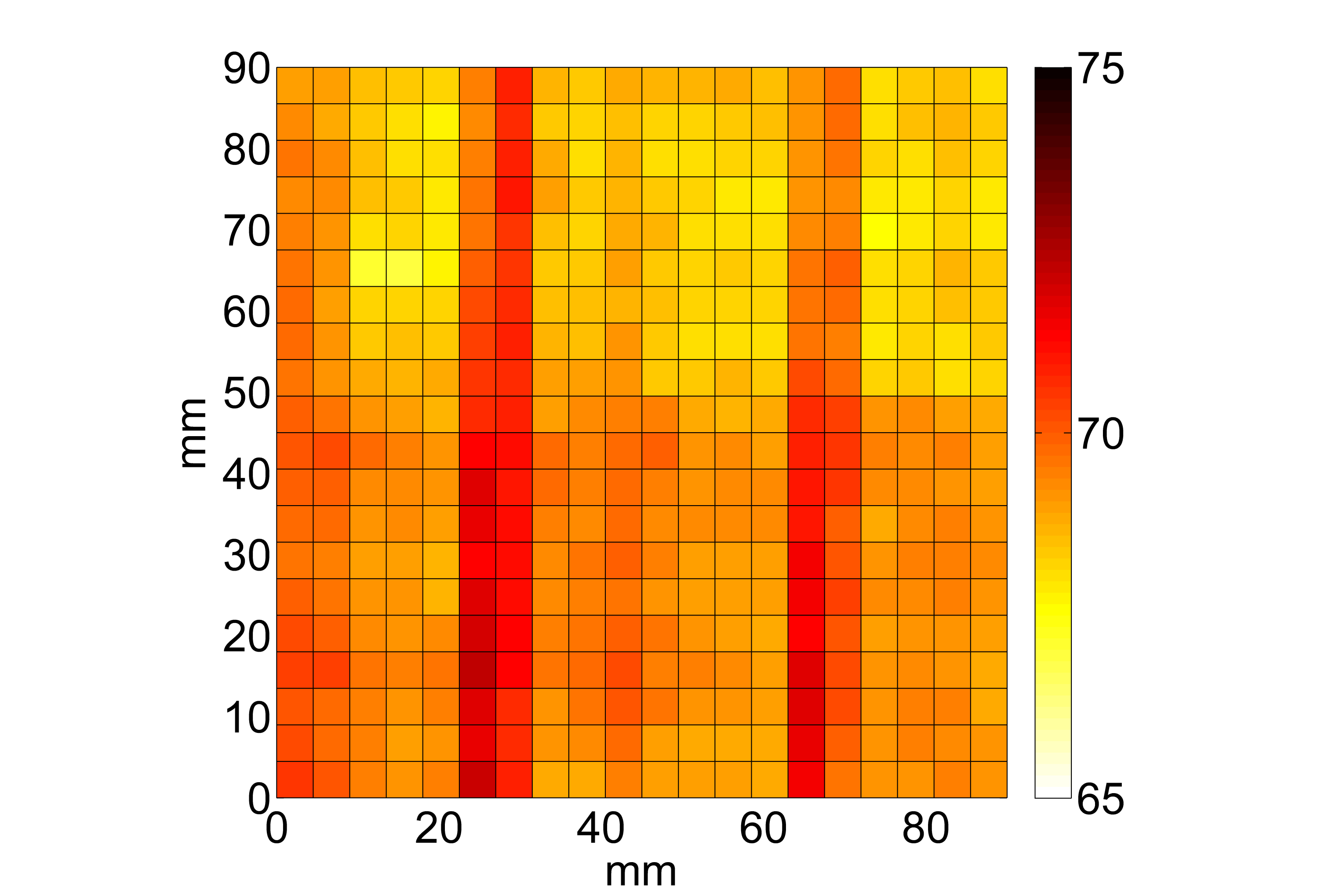}
\caption{2d histogram of the measured diameters of the outer holes on the bottom side of foil 2.\label{outerbot}}
\end{figure}

\begin{figure}[htp]\centering
\includegraphics[width=0.47\textwidth]{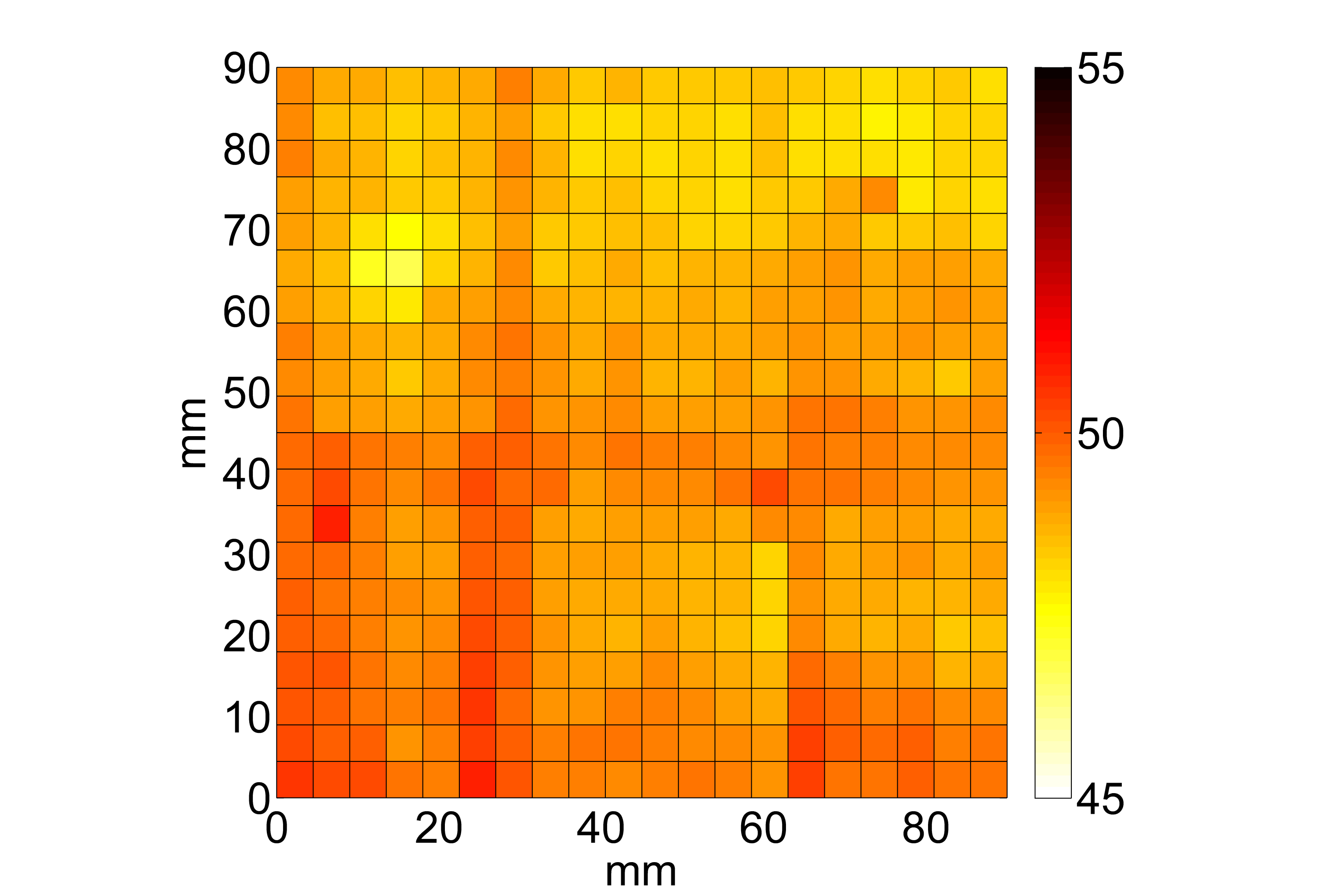}
\caption{2d histogram of the measured diameters of the inner holes of foil 2.\label{inner}}
\end{figure}

For completeness, the corresponding diameter histograms are shown in Figures ~\ref{outertop} - ~\ref{inner}. The color map of the histograms is inverted to emphasize the connection of the hole size and the measured gain.


\section{Conclusions}

Analysis software was developed for the QA of GEM foils, which are essential components in GEM detectors and are widely being developed for future HEP and nuclear physics experiments.

The software was tested by comparing measurements made from both sides of a set of 5 GEM foils. The inner and outer diameters of the holes, the pitch of the hole pattern and the difference between the length of the axes of an ellipse fitted on the hole boundary were used as the measured parameters. Measurement uncertainties for the parameters, with $2\sigma$ confidence level, were found to be 1.05~$\mu$m, 0.65~$\mu$m and 1.24~$\mu$m, respectively. Relative differences of GEM hole diameters within a single scan can be resolved with $2\sigma$ uncertainty of 0.80~$\mu$m.

The analysis software can be used to accurately characterize the outer and the inner hole diameters and the pitch of the GEM foils, the properties relevant to the performance of the foils. Furthermore, the ellipticity of the holes can be estimated, potentially indicating problems with the foil manufacturing process, such as mask alignment errors.

It was shown that a qualitative estimation of the behavior of the local variation in gain across the GEM foil can be made based on the measured sizes of the outer and inner holes. Due to the multiplicative nature of the gain in multiple foil detectors, the total variation can become locally significant and should be taken into account in QA if gain homogeneity is of importance. A more descriptive model of the gain, using FEM-based electric field maps and gas multiplication simulations with Garfield for bi-conical and single-mask holes is subject for further studies.

\section*{Acknowledgments}
The authors would like to thank Eraldo Oliveri for his numerous advices and help. They also thank the RD51 collaboration for using the Gaseous Detector Laboratory facilities at CERN.

\clearpage

\section*{References}

\bibliography{opticalQA4}


\end{document}